\documentclass[structabstract]{aa}  
\usepackage{graphicx}
\usepackage{natbib}
\usepackage{amsmath}
\usepackage{rotating}
\usepackage{lscape}
\usepackage{color}
\usepackage{latexsym}
\usepackage{amsfonts}
\usepackage{multirow}
\usepackage{rotating}
\usepackage{graphicx}
\usepackage{epsfig}
\usepackage{amssymb,amsbsy}
\bibpunct{(}{)}{;}{a}{}{,}
\newcommand{\ie}{$i.e.,\;$}
\newcommand{\eg}{$e.g.,\;$}
\newcommand{\viz}{$viz.,\;$}
\newcommand{\cf}{$cf.,\;$}

\usepackage{txfonts}
\begin{document}
    \title{Multiwavelength characterization of faint Ultra Steep Spectrum radio sources : A search for high-redshift radio galaxies 
}

   \author{Veeresh Singh\inst{1}\fnmsep\thanks{veeresh.singh@ias.u-psud.fr}, Alexandre Beelen\inst{1}, Yogesh Wadadekar\inst{2}, 
Sandeep Sirothia\inst{2}, Ishwara-Chandra C.H.\inst{2}, Aritra Basu\inst{2}, Alain Omont\inst{3}, Kim McAlpine\inst{4}, 
R.~J.~Ivison\inst{5,6}, Seb Oliver\inst{7}, Duncan Farrah\inst{7,8}, \and Mark Lacy\inst{9}
}
   \institute{
Institut d'Astrophysique Spatiale, B$\hat{\rm a}$t. 121, Universit{\'e} Paris-Sud, 91405 Orsay Cedex, France
\and  
National Centre for Radio Astrophysics, TIFR, Post Bag 3, Ganeshkhind, Pune 411007, India
\and 
UPMC Univ Paris 06 and CNRS, UMR 7095, Institut d'Astrophysique de Paris, F-75014, Paris, France
\and Department of Physics, University of the Western Cape, Private Bag X17, Bellville 7537, South Africa
\and 
European Southern Observatory, Karl Schwarzschild Strasse 2, D-85748 Garching, Germany
\and Institute for Astronomy, University of Edinburgh, Blackford Hill, Edinburgh EH9 3HJ, UK
\and Astronomy Centre, Department of Physics and Astronomy, University of Sussex, Brighton, BN1 9QH, UK
\and Department of Physics, Virginia Tech, Blacksburg, VA 24061, USA
\and National Radio Astronomy Observatory, 520 Edgemont Road, Charlottesville, VA 22903, USA
}

\date{}

 
  \abstract
{Ultra Steep Spectrum (USS) radio sources are one of the efficient tracers of powerful High-$z$ Radio Galaxies (H$z$RGs). 
In contrast to searches for powerful H$z$RGs from radio surveys of moderate depths, fainter USS samples derived from deeper radio surveys 
can be useful in finding H$z$RGs at even higher redshifts and in unveiling a population of obscured weaker radio$-$loud AGN at moderate redshifts.   
}
{Using our 325 MHz GMRT observations (5$\sigma$ $\sim$ 800 $\mu$Jy) and 1.4 GHz VLA observations (5$\sigma$ $\sim$ 80 $-$ 100 $\mu$Jy) 
available in two subfields ({\viz} VLA-VIMOS VLT Deep Survey (VLA-VVDS) and Subaru X-ray Deep Field (SXDF)) 
of the XMM-LSS field, we derive a large sample of 160 faint USS radio sources and characterize their nature. 
}
{The optical, IR counterparts of our USS sample sources are searched using existing deep surveys, at respective wavelengths.
We attempt to unveil the nature of our faint USS sources using diagnostic techniques based on mid-IR colors, flux ratios of radio to mid-IR, and radio 
luminosities.   
}
{Redshift estimates are available for 86/116 ($\sim$ 74$\%$) USS sources in the VLA-VVDS field and for 39/44 ($\sim$ 87$\%$) USS sources in the 
SXDF fields with median values ($z_{\rm median}$) $\sim$ 1.18 and $\sim$ 1.57, 
which are higher than that for non-USS radio sources ($z_{\rm median~non-USS}$ $\sim$ 0.99 and $\sim$ 0.96), in the two subfields, respectively. 
The MIR color-color diagnostic and radio luminosities are consistent with a majority of our USS sample sources at higher redshifts ($z > 0.5$) being AGN.
The flux ratio of radio to mid-IR (S$_{\rm 1.4~GHz}$/S$_{\rm 3.6~{\mu}m}$) versus redshift diagnostic plot suggests that more than half of our 
USS sample sources distributed over $z$ ${\sim}$ 0.5 to 3.8 are likely to be hosted in obscured environments. 
A significant fraction ($\sim$ 26$\%$ in the VLA-VVDS and $\sim$ 13$\%$ in the SXDF) of our USS sources without redshift estimates mostly remain 
unidentified in the existing optical, IR surveys, and exhibit high radio to mid-IR flux ratio limits similar to H$z$RGs, and thus, can be considered 
as potential H$z$RG candidates.           
}    
{Our study shows that the criterion of ultra steep spectral index remains a reasonably efficient method to select high-$z$ sources even at sub-mJy flux densities. 
In addition to powerful H$z$RG candidates, our faint USS sample also contain population of weaker radio$-$loud AGNs potentially hosted in 
obscured environments.  
}


   \maketitle              
%

\section{Introduction}

High-$z$ radio galaxies (H$z$RGs) are found to be hosted in massive intensely star forming galaxies which contain large reservoirs of dust and gas 
({\eg}\cite{Eales96,Jarvis01a,Willott03,DeBreuck05,Klamer05,Seymour07}).
Host galaxies of H$z$RGs are believed to be the progenitors of massive elliptical galaxies present in the local universe, 
as the powerful radio galaxies in the local universe are hosted in massive ellipticals \citep{Best98,McLure04}. 
H$z$RGs are also often found to be associated with over-densities {\ie} proto-clusters and clusters of galaxies at redshifts ($z$) $\sim$ 2 - 5 
({\eg}\cite{Stevens03,Kodama07,Venemans07,Galametz12}). 
Therefore, identification and study of H$z$RGs helps us to better understand the formation and evolution of galaxies at higher redshifts and in 
dense environments.  
The correlation between the steepness of the radio spectrum and cosmological redshift ({\ie}$z-{\alpha}$ correlation) has been exploited as one of 
the successful tracers to find H$z$RGs \citep{Roettgering94,Chambers96,DeBreuck2000,DeBreuck02,Klamer06,Ishwara-Chandra10, Ker12}. 
In fact, most of the radio galaxies known at $z > 3.5$ have been found using the Ultra Steep Spectrum (USS) criterion 
\citep{Blundell98,DeBreuck98,DeBreuck2000,Jarvis01a,Jarvis01b,DeBreuck02a,Jarvis04,Cruz06,Miley08}.
The causal connection between the steepness of radio spectral index and redshift is not well understood.
The radio spectral index may become steeper at high redshift possibly due to an increased spectral curvature 
with redshift and the redshifting of a concave radio spectrum to lower radio frequencies ({\eg}\cite{Krolik91}). 
The steepening of radio spectrum may also be caused if radio jets expand in denser environments, a scenario which could be more viable 
in proto-cluster environments in the distant Universe \citep{Klamer06,Bryant09,Bornancini10}. 
In general, a large fraction of H$z$RGs are found in samples of USS ($\alpha$ $\leq$ -1.0 with S$_{\nu}$ $\propto$ ${\nu}^{\alpha}$) 
radio sources, however, an USS can not be guaranteed as a high redshift source and vice-versa 
({\eg}\cite{Waddington99,Jarvis09}).
Since radio emission does not suffer from dust absorption, the selection of H$z$RGs 
at radio frequency yields an optically unbiased sample. \par
Until recently, most studies on H$z$RGs using USS samples were limited to brighter sources ({\eg}S$_{\rm 1.4~GHz}$ $\geq$ 10 mJy) derived from 
shallow or moderately deep, wide area radio surveys ({\eg}\cite{DeBreuck02,DeBreuck04,Broderick07,Bryant09,Bornancini10}). 
This raises the question whether faint USS sources represent a population of powerful radio galaxies at even higher redshifts 
or a population of low-power AGNs at moderate redshifts or a mixed population of both classes.    
Low frequency radio observations are more advantageous in finding faint USS sources as their flux density is higher at low-frequency 
due to their steeper spectral index. Sensitive low frequency radio observations with the Giant Metrewave Radio Telescope (GMRT) have 
become useful to search and study USS sources with S$_{\rm 1.4~GHz}$ down to submJy level ({\eg}\cite{Bondi07,Ibar09,Afonso11}).
Furthermore, it is interesting to study faint USS sources down to submJy level, as the radio population at submJy level appears to be different than 
that at brighter end (above few mJy) and an increasingly large contribution from the evolving star$-$forming galaxy population is believed to be 
present at submJy level \citep{Afonso05,Simpson06,Smolcic08}. \par
\begin{figure}
\centering
\includegraphics[angle=0,width=8.8cm,height=8.2cm]{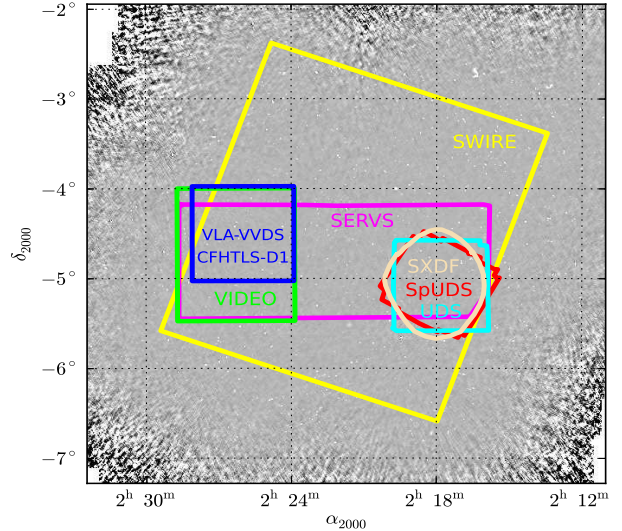}
\caption{Footprints of VLA-VVDS (B03 field; in blue), SXDF (S06 field; in wheat), VIDEO (in green), SERVS (in magenta), UDS (in cyan), 
SpUDS (in red) and SWIRE (in yellow) fields overplotted on our 325 MHz GMRT image. CFHTLS-D1 covers the same area as VLA-VVDS.}
\label{fig:footprints}
\end{figure}

In this paper, we study the nature of faint USS sources derived from our 325 MHz low-frequency GMRT observations and 1.4 GHz VLA 
observations over the two subfields {\viz}the VLA$-$VIMOS VLT Deep Survey (VLA$-$VVDS) field \citep{Bondi03} and 
the Subaru X-ray Deep Field (SXDF) \citep{Simpson06} in the XMM-LSS field. 
Hereafter, we refer to \cite{Bondi03} as `B03' and to \cite{Simpson06} as `S06'. 
The sky coverages in 1.4 GHz radio observations of B03 ({\ie}VLA$-$VVDS field) and of S06 ({\ie}SXDF field) are named 
as the `B03 field' and `S06 field', respectively. 
Figure~\ref{fig:footprints} shows the footprints of \cite{Bondi03,Bondi07} and \cite{Simpson06} 1.4 GHz observations 
plotted over our 325 MHz image. 
We present our {analysis} on the two subfields separately as the available multiwavelength data in the two subfields come from 
different surveys and are of different sensitivities.
In Section 2, we discuss the radio observations in the two subfields and our USS sample selection. 
The optical, near-IR and mid-IR identification of our USS sources is discussed in Section 3. 
The redshift distributions of our USS sources are discussed in Section 4. 
The mid-IR color-color diagnostics and the properties of flux ratios of radio to mid-IR fluxes are discussed in Section 5. 
In Section 6, we discuss the radio luminosity distributions of our USS sources.
Section 7 is devoted to examining the $K-z$ relation for our faint USS sources. 
In Section 8, we discuss the efficiency of the USS technique in selecting high-$z$ sources at faint flux densities. 
We present the conclusions of our study in Section 9. Our full USS sample is given in the Appendix Table~\ref{tab:USSSample}.
\\ 
We adopt cosmological parameters H$_{\rm 0}$ = 71 km s$^{-1}$ Mpc$^{-1}$, ${\Omega}_{\rm M}$ = 0.27 and ${\Omega}_{\Lambda}$ = 0.73 throughout this paper. 
All the quoted magnitudes are in the AB system unless stated otherwise.

\section{USS sample selection}
\subsection{325 MHz GMRT observations of the XMM-LSS}
We obtained 325 MHz GMRT observations of the XMM-LSS field over sky area of $\sim$ 12 deg$^{2}$ 
with synthesized beamsize $\sim$ 10$^{{\prime}{\prime}}$.2 $\times$ 7$^{{\prime}{\prime}}$.9. 
In the mosaiced 325 MHz GMRT image the average noise rms is $\sim$ 160 $\mu$Jy, while in the central region the 
average noise-rms reaches down to $\sim$ 120 $\mu$Jy. 
Our 325 MHz observations are one of the deepest low-frequency surveys over such a wide sky area and detect $\sim$ 2553 / 3304 radio sources 
at $\geq$ 5.0$\sigma$ with noise rms cut-off $\leq$ 200 / 300 $\mu$Jy. 
Since the local noise rms varies with distance from the phase center and also in the vicinity of bright sources, 
the rms map was used for source extraction and this approach helped to minimize the detection of spurious sources.
We only consider sources with peak source brightness greater than 5 times the local rms noise value.
The source position (right ascension and declination) is determined as the flux-density weighted centroid of all the emission enclosed 
within the 3$\sigma$ contour. 
The typical error in the positions of the sources is about 1.4 arcsec and is estimated using the formalism outlined by \cite{Condon98}. 
The procedures opted for the data reduction and  source extraction are similar to the 325 MHz GMRT observations of ELAIS-N1 presented in \cite{Sirothia09}.
The details of our radio observations, data reduction, and source catalog of the XMM-LSS field will be presented in 
Sirothia et al. (2014; in preparation). 
We note that our 325 MHz observations are $\sim$ 5 times deeper than the previous 
325 MHz observations of the XMM-LSS field (\cite{Tasse06,Cohen03}), and result in similar manifold increase in the source density. 
Also, our 325 MHz observations are $\sim$ 3 times more sensitive 
(assuming typical spectral index for radio sources $\alpha$ $\simeq$ -0.7) than the existing 610 MHz observations in the XMM-LSS (\cite{Tasse07}).
\subsection{Other radio observations in the XMM-LSS field}
The XMM-LSS field has been observed at different radio frequencies with varying sensitivities and sky area coverages 
({\eg}\cite{Bondi03,Cohen03,Bondi07,Simpson06,Tasse06,Tasse07}). 
Among the deep surveys, there are 1.4 GHz and 610 MHz observations of 1.0 deg$^{2}$ in the VVDS field \citep{Bondi03,Bondi07} and 
1.4 GHz observations of 1.3 deg$^{2}$ in the SXDF fields \citep{Simpson06}.
The 1.4 GHz VLA observations of 1.0 deg$^{2}$ in the VLA-VVDS field detect total $\sim$ 1054 radio sources above 5$\sigma$ limit ($\sim$ 80 $\mu$Jy)   
with resolution of $\sim$ 6.0$^{{\prime}{\prime}}$ \citep{Bondi03}.
The 610 MHz GMRT observations of the same area in the VLA-VVDS field detect total $\sim$ 512 radio sources above 5$\sigma$ limit 
($\sim$ 250 $\mu$Jy) with resolution of $\sim$ 6.0$^{{\prime}{\prime}}$ \citep{Bondi07}.
\cite{Simpson06} present 1.4 GHz VLA observations of $\sim$ 1.3 deg$^{2}$ in the SXDF field and detect $\sim$ 512 sources over 
central $\sim$ 0.8 deg$^{2}$ above 5$\sigma$ detection limit ($\sim$ 100 $\mu$Jy).
\subsection{Cross-matching of 325 MHz sources and 1.4 GHz sources}
We cross-match 325 MHz GMRT sources with 1.4 GHz VLA sources in the B03 and the S06 subfields and select our sample of USS 
sources based on 325 MHz to 1.4 GHz spectral index.
To cross-match 325 MHz sources with 1.4 GHz sources we follow the method proposed by \cite{Sirothia09}.
We identify 1.4 GHz counterparts of 325 MHz sources by using a search radius of 7.5 arcsec for unresolved sources and a larger search radius equal 
to the sum of half of the angular size and 7.5 arcsec for resolved sources.  
The value of search radius is approximately equal to the sum of the half power synthesized beamwidths at 1.4 GHz and 325 MHz. 
We checked with increasing search radii from 7${^{\prime}}{^{\prime}}$.5 to 10${^{\prime}}{^{\prime}}$ and 15${^{\prime}}{^{\prime}}$, and found 
that the number of unresolved cross-matched sources remains nearly same. 
Since the radio source density is low {\ie}only 1054 sources detected at 1.4 GHz over 1.0 deg$^{-2}$, the chance coincidence in our cross-matching 
of 325 MHz sources to 1.4 GHz radio sources is rather small {\ie}0.14$\%$.
The cross-matching of 325 MHz and 1.4 GHz radio source catalogs yields a total of 338 and 190 cross-matched sources in the B03 and the S06 
subfields, respectively ({\cf}~Table\ref{table:USSNumbers}). 
There are a large number of faint 1.4 GHz sources without 325 MHz counterparts and this can be understood as the 1.4 GHz observations are 
much deeper ($\sim$ 80 - 100 $\mu$Jy at 5$\sigma$ level) compared to the 325 MHz observations. 
However, the 5$\sigma$ detection limit ($\sim$ 800 $\mu$Jy) of our 325 MHz observations corresponds to 
$\sim$ 288 $\mu$Jy at 1.4 GHz, assuming typical spectral index for radio sources ($\alpha$) $\sim$ -0.7. 
Also, there are a few 325 MHz detected radio sources that are not detected in the 1.4 GHz observations at $\geq$ 5.0$\sigma$.  
These sources can be explained if they have ultra steep spectral index ($\alpha{_{\rm 325~MHz}^{\rm 1.4~GHz}}$ $\leq$ -1.3).
We discuss these sources in the next sub-section.

\subsection{325 MHz $-$ 1.4 GHz radio spectral index}
\begin{figure}
\includegraphics[angle=0,width=9.0cm,height=8.0cm]{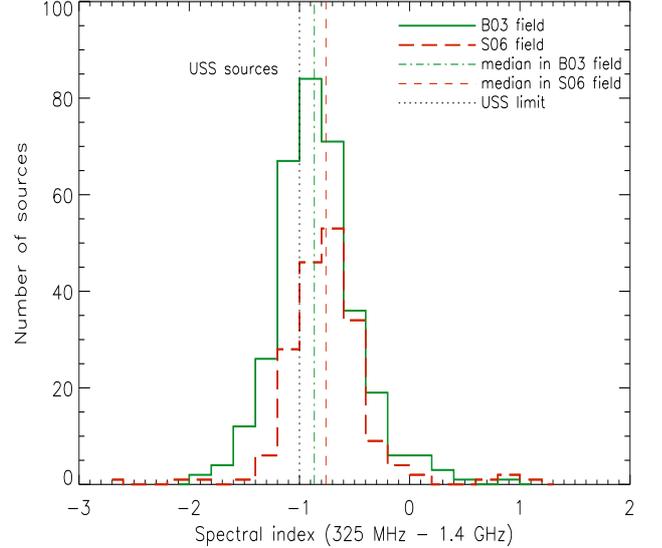}
\caption{Histogram of 325 MHz to 1.4 GHz spectral index (${\alpha}_{\rm 325~MHz}^{\rm 1.4~GHz}$). Green solid line histogram represents sources in the B03 field while 
red dashed line histogram represents sources in the S06 field. Median spectral indices in the B03 field 
(${\alpha}_{\rm 325~MHz,~median}^{\rm 1.4~GHz}$ $\sim$ -0.87) and in the S06 field (${\alpha}_{\rm 325~MHz,~median}^{\rm 1.4~GHz}$ $\sim$ -0.76) are 
represented by vertical green dashed-dotted and red dashed lines, respectively. USS limit (${\alpha}_{\rm 325~MHz}^{\rm 1.4~GHz}$ $\leq$ -1) 
is represented by the vertical dotted line.  
}
\label{fig:SpInHist}
\end{figure}

\begin{figure}
\includegraphics[angle=0,width=8.8cm,height=7.6cm]{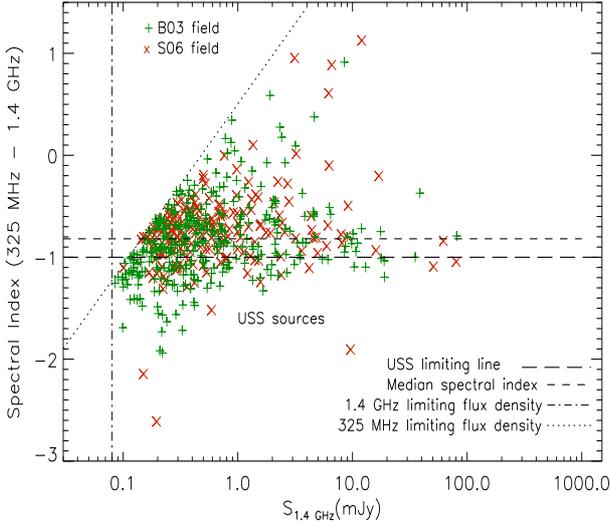}
\caption{Spectral index (${\alpha}_{\rm 325~MHz}^{\rm 1.4~GHz}$) versus 1.4 GHz flux density plot. 
Plus (`+') and cross (`${\times}$') symbols represent USS sources in the B03 and in the S06 fields, respectively. 
The dashed line represents the median spectral index value (${\alpha}_{\rm 325~MHz}^{\rm 1.4~GHz}$ $\sim$ -0.83) for the full sample. 
The flux density limits at 325~MHz and 1.4~GHz are represented by dotted and long-dashed lines, respectively. 
}
\label{fig:FDVsSpIn}
\end{figure}

We estimate radio spectral index ($\alpha$, where S$_{\nu}$ $\propto$ $\nu$$^{\alpha}$) for all the sources that are detected at both 325 MHz 
and 1.4 GHz frequencies. 
Figure~\ref{fig:SpInHist} shows the histograms of spectral index of cross-matched sources for both the B03 and the S06 fields.
The median values of the spectral index distributions ($\alpha{_{\rm 325~MHz}^{\rm 1.4~GHz}}$) are -0.86 (standard deviation $\sim$ 0.38) 
and -0.76 (standard deviation $\sim$ 0.40) in the B03 and the S06 fields, respectively. 
The higher median spectral index in the B03 field is possibly due to the deeper 1.4 GHz source catalog {\ie}faint 1.4 GHz sources with 
steeper spectral index are favored to be detected at 325~MHz. Figure~\ref{fig:FDVsSpIn} shows the 1.4 GHz flux density versus spectral index 
($\alpha{_{\rm 325~MHz}^{\rm 1.4~GHz}}$) plot. 
The differing sensitivities at the two frequencies result in a bias against flat spectral index sources {\ie} faint 1.4 GHz sources 
with relatively flat spectral index have corresponding 325 MHz flux density below the detection limit of less sensitive 325 MHz observations.  
The large number of sources lying along the 325 MHz flux density limit line in the spectral index versus flux density 
plot reflects the fact that 1.4 GHz observations are deeper than 325 MHz observations.   
\subsection{USS Sample}
In the literature there is no uniform definition for a USS source and different studies have used different frequencies and different spectral index 
thresholds {\eg}${\alpha}^{\rm 4.85~GHz}_{\rm 151~MHz}$ $\leq$ -0.981 \citep{Blundell98}, 
${\alpha}^{\rm 325~MHz}_{\rm 74~MHz}$ $\leq$ -1.2 \citep{Cohen04}, ${\alpha}^{\rm 843~MHz}_{\rm 408~MHz}$ $\leq$ -1.3 \citep{DeBreuck04},
 ${\alpha}^{\rm 1.4~GHz}_{\rm 151~MHz}$ $\leq$ -1.0 \citep{Cruz06}, ${\alpha}^{\rm 843~MHz}_{\rm 408~MHz}$ $\leq$ -1.0 \citep{Broderick07} and 
${\alpha}^{\rm 1.4~GHz}_{\rm 150~MHz}$ $\leq$ -1.0 \citep{Ishwara-Chandra10}.
To select our sample of USS sources we use spectral index cut-off  ${\alpha}^{\rm 1.4~GHz}_{\rm 325~MHz}$ $\leq$ -1.0 
(spectra steeper than -1.0). 
The spectral index may change with frequency due to spectral curvature \citep{Bornancini07}, although majority of H$z$RGs show linear 
spectra over a large frequency range \citep{Klamer06}. 
Thus, a higher cut-off in the spectral index at 325 MHz will translate into even 
higher cut-off at the rest frame, if a source exhibits spectral steepening at higher frequencies. 
Furthermore, at fainter flux densities, the less-luminous radio sources can have marginally flatter spectra due to observed correlation between 
the radio power and the spectral index {\ie}the P$-$~$\alpha$ relation \citep{Mangalam95,Blundell99}.
Since we are studying faint USS sources to identify H$z$RGs there is a possibility that a large fraction of H$z$RGs may be missed 
if we adopt a very steep spectral index cut-off 
({\eg}$\alpha$ $\leq$ -1.3). Moreover, if we happen to pick up low redshift sources in our USS sample by using a less steep spectral index cut-off, 
these sources are likely to have optical counterparts and redshift estimates, and therefore can be identified and eliminated. 
Using the spectral index $\alpha{_{\rm 325~MHz}^{\rm 1.4~GHz}}$ $\leq$ -1.0 for a source to be classified as Ultra Steep Spectrum (USS) 
source in the 325 MHz - 1.4 GHz cross-matched catalogs, we obtain 111 and 39 USS sources in the B03 and S06 fields, respectively 
({\cf}~Table\ref{table:USSNumbers}). \\
There are 5 radio sources in each subfield that are detected in 325 MHz at $\geq$ 5$\sigma$ but do not have 1.4 GHz counterpart 
at $\geq$~5$\sigma$ flux limit.
These sources are potential faint USS sources as due to very steep spectral index they are detected above 5$\sigma$ at 325 MHz but 
fall below 5$\sigma$ detection at 1.4 GHz.
To find the 1.4 GHz counterparts of such sources we inspected 1.4 GHz images and find that all sources are 
detected between 3$\sigma$ to 5$\sigma$ level. We obtained their 1.4 flux densities by fitting the source with an elliptical Gaussian using the task `JMFIT' 
in `AIPS'\footnote{http://www.aips.nrao.edu}. 
It turns out that some of these sources are marginally resolved with peak flux density below 5$\sigma$ while total 
flux density is above 5$\sigma$. 
Thus, the resultant spectral index is not as steep as expected from the 5$\sigma$ detection flux limit at 1.4 GHz.
The addition of these USS sources (detected above 5$\sigma$ at 325 MHz but falling below 5$\sigma$ at 1.4 GHz) to those detected at $\geq$ 5$\sigma$ in both frequencies result, 
in total, 116 and 44 USS sources in the B03 and the S06 fields, respectively, and a full sample of 160 USS sources 
({\cf}Table~\ref{table:USSNumbers}).

The flux density measurement errors give rise to uncertainties in spectral indices and this could result in scattering of some non-USS sources into 
the USS sample and vice-versa. In order to statistically quantify the contamination of non-USS sources into the USS sample, we consider 
spectral index distribution of 325 MHz selected sources described by a normal distribution of 
$\bar{\alpha}{_{\rm 325~MHz}^{\rm 1.4~GHz}}$ $\pm$ ${\sigma}_{\alpha}$ $=$ -0.82 $\pm$ 0.39, and the 
distributions of errors on spectral indices described by a normal distribution 
of ${\Delta}{\alpha}$ $\pm$ ${\sigma}_{{\Delta}{\alpha}}$ $=$ 0.08 $\pm$ 0.05. 
As our spectral index cut-off for USS sources $\alpha{_{\rm 325~MHz}^{\rm 1.4~GHz}}$ $=$ -1.0 lies at steep tail of 
the spectral index distribution, more number of non-USS sources $\alpha{_{\rm 325~MHz}^{\rm 1.4~GHz}}$ $>$ -1.0 are expected to 
scatter into the USS sample than the USS sources scatter to non-USS regime. 
Using the median uncertainty of spectral indices and a normal distribution for spectral indices we find that 48 non-USS sources 
with observed spectral index $\alpha{_{\rm 325~MHz}^{\rm 1.4~GHz}}$ $>$ -1.0 may have intrinsic spectral index 
$\alpha{_{\rm 325~MHz}^{\rm 1.4~GHz}}$ $\leq$ -1.0, while 43 USS sources with observed spectral index 
$\alpha{_{\rm 325~MHz}^{\rm 1.4~GHz}}$ $\leq$ -1.0 may have intrinsic spectral index 
$\alpha{_{\rm 325~MHz}^{\rm 1.4~GHz}}$ $>$ -1.0. 
This indicates that the contamination by non-USS sources in our sample can be as large as 48/160 $\sim$ 30$\%$. 
The contamination by intrinsically non-USS sources is likely to result in the increase of low-$z$ sources in our USS sample.

\begin{table} 
\centering
\caption{Radio sources}
\begin{minipage}{140mm}
\begin{tabular} {ccc}
\hline
Total no. of sources  & \multicolumn{2}{c}{Field} \\
                   & B03 & S06 \\ \hline
detected at 1.4 GHz ($\geq$ 5$\sigma$) &  1054         & 512          \\
detected at 325 MHz ($\geq$ 5$\sigma$) &  343          & 195       \\
cross-matched sources & 338         & 190           \\
                                          &                  &                 \\
USS sources (${\rm {\alpha}_{325~MHz}^{1.4~GHz}}$ $\leq$ -1.0)  &    &    \\ 
S$_{\rm 325~MHz}$ $\geq$ 5$\sigma$ and S$_{\rm 1.4~GHz}$ $\geq$ 5$\sigma$ &  111  &  39 \\ 
                                         &                  &                 \\
USS sources (${\rm {\alpha}_{325~MHz}^{1.4~GHz}}$ $\leq$ -1.0) &    &   \\ 
S$_{\rm 325~MHz}$ $\geq$ 5$\sigma$ and S$_{\rm 1.4~GHz}$ $\sim$ 3$\sigma$ $-$ 5$\sigma$ &  5  & 5 \\ 
                                          &                  &                 \\
All USS sources (${\rm {\alpha}_{325~MHz}^{1.4~GHz}}$ $\leq$ -1.0) &  116  &  44  \\ \hline
\end{tabular} 
\label{table:USSNumbers} 
\vspace {0.1cm} \\
\end{minipage}
\end{table}

\subsection{Comparison with 610 MHz - 1.4 GHz USS sample}
\cite{Bondi07} present a sample of 58 faint USS sources (${\alpha}_{\rm 610~MHz}^{\rm 1.4~GHz}$ $\leq$ -1.3) using deep 1.4 GHz 
(5$\sigma$ $\sim$ 80 $\mu$Jy) and 610 MHz (5$\sigma$ $\sim$ 250 $\mu$Jy) observations of 1.0 deg$^{-2}$ in the VLA-VVDS field. 
39/58 of these USS sources have 1.4 GHz detection at $\geq$~5$\sigma$ and 610 MHz detection at $\geq$~3$\sigma$, while rest of the 
19/58 USS sources have 610 MHz detection at $\geq$~5$\sigma$ but 1.4 GHz detection is between 3$\sigma$ to 5$\sigma$.
We derive our USS sample (${\alpha}_{\rm 325~MHz}^{\rm 1.4~GHz}$ $\leq$ -1.0) in the same field using low frequency 325 MHz observations 
and 1.4 GHz observations. 
We find that only 11 USS sources are common to our USS sample (${\alpha}_{\rm 325~MHz}^{\rm 1.4~GHz}$ $\leq$ -1.0) 
and the USS sample of \cite{Bondi07} (${\alpha}_{\rm 610~MHz}^{\rm 1.4~GHz}$ $\leq$ -1.3). The mismatch could be attributed to different flux limits as we have considered only those sources that 
are detected at $\geq$ 5$\sigma$ at both the 1.4 GHz and 325 MHz. 
\cite{Bondi07} cautioned that all their 58 USS candidates are weak radio sources 
({\ie}50 $\mu$Jy $\leq$~S$_{\rm 1.4~GHz}$~$\leq$ 327 $\mu$Jy, with the median S$_{\rm 1.4~GHz}$ $\sim$ 90 $\mu$Jy), 
and therefore, errors in the total flux density determination can be relatively large, yielding to a less secure spectral index value. 
Since USS sources are faint and unresolved, we used peak flux densities and find that 
22/58 USS sources have extrapolated 325 MHz flux density below the detection limit 
of our GMRT observations ({\ie}S$_{\rm 325~MHz}$ $<$ 0.80 $\mu$Jy). 
The non-detection of rest of the 25/58 sources at 325 MHz can be explained if these sources exhibit spectral turnover between 325 MHz to 610 MHz, 
or if there is large uncertainty associated with 610 MHz - 1.4 GHz spectral index (${\alpha}_{\rm 610~MHz}^{\rm 1.4~GHz}$). 
The possibility of some of the sources being Giga-hertz Peaked Sources (GPS) like or affected by variability cannot be ruled out.   
For rest of the 105 USS sources (${\alpha}_{\rm 325~MHz}^{\rm 1.4~GHz}$ $\leq$ -1.0) of our sample, 80, 16, and 9 sources have 610 MHz detection 
at $\geq$ 5$\sigma$, 3$\sigma$ $-$ 5$\sigma$, and $<$3$\sigma$, respectively. Majority of our USS sources (${\alpha}_{\rm 325~MHz}^{\rm 1.4~GHz}$ $\leq$ -1.0) 
have ${\alpha}_{\rm 610~MHz}^{\rm 1.4~GHz}$ $\sim$ -1.3 to -0.7, which is consistent within uncertainties.

\section{The optical, near-IR and mid-IR counterparts of USS sources}
To characterize the nature of our USS radio sources we study the properties of their counterparts in different bands at optical and IR wavelengths. 
\subsection{The optical, near-IR and mid-IR data}
{\bf The B03 field} : 
To find the optical counterparts of our USS sources, we use VLT VIMOS Deep Survey (VVDS${\footnote{http://cesam.oamp.fr/vvdsproject//index.html}}$) 
and Canada-France-Hawaii Telescope Legacy Survey (CFHTLS${\footnote{http://www.cfht.hawaii.edu/Science/CFHLS/}}$) D1 photometric data. 
\cite{Ciliegi05} present optical identification of 1.4 GHz radio sources using VVDS photometric data in B, V, R and I bands.  
In near-IR, we use VISTA Deep Extragalactic Observations (VIDEO; \cite{Jarvis13}) survey which provides photometric observations 
in Z, Y, J, H and Ks bands and covers full 1.0 deg$^{-2}$ of the B03 field. 
\cite{McAlpine13} cross-matched 1.4 GHz radio sources to the K-band VIDEO data and also used CFHTLS-D1 photometric data 
in u$^{\star}$, g$^{\prime}$, r$^{\prime}$, i$^{\prime}$ and z$^{\prime}$ bands along with VIDEO photometric data to obtain 
photometric redshift estimates of 1.4 GHz radio sources.   
To find mid-IR counterparts we 
use Spitzer Extragalactic Representative Volume Survey (SERVS) data \citep{Mauduit12}. 
SERVS is a medium deep survey at 3.6 and 4.5 $\mu$m and has partial overlap of $\sim$ 0.82 deg$^{-2}$ with 
the B03 field ({\cf}Figure~\ref{fig:footprints}). \\
{\bf The S06 field} : \cite{Simpson06} present optical identifications of 1.4 GHz radio sources 
using the Subaru/Suprime-Cam observations in B, V, R, i$^{\prime}$, z$^{\prime}$ bands. 
To find the optical counterparts of our USS sources we use optical radio cross-matched catalog of \cite{Simpson06}.  
In near-IR, we use the Ultra Deep Survey${\footnote{http://www.nottingham.ac.uk/astronomy/UDS}}$ (UDS) DR8 from the UKIRT Infrared Deep Sky Survey 
(UKIDSS, \cite{Lawrence07}) which has $\sim$ 0.63 deg$^{-2}$ of overlap with the S06 field. 
The mid-IR counterparts are found using the Spitzer Public Legacy Survey of the UKIDSS Ultra Deep Survey 
(SpUDS$\footnote{http://irsa.ipac.caltech.edu/data/SPITZER/SpUDS/}$) \citep{Dunlop07} which is carried out with all four IRAC bands 
(3.6, 4.5, 5.8 and 8.0 $\mu$m) and one MIPS band (24 $\mu$m).  

\subsection{The optical, near-IR and mid-IR identification rates}
\begin{figure}
\centering
\includegraphics[angle=0,width=8.8cm,height=7.6cm]{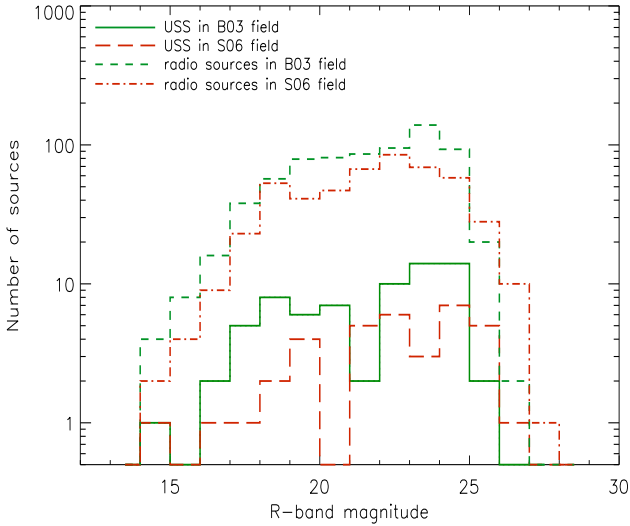}
\caption{Histograms of R-band magnitudes of the USS sources and of the full 1.4 GHz radio population in the B03 and the S06 field. 
Histograms of USS sources are shown by green solid lines and red long dashed lines for the B03 and the S06 fields, respectively. 
While green dashed and red dashed-dotted lines represent histograms for the full 1.4 GHz radio population in the B03 and the S06 fields, respectively.}
\label{fig:RMagHist}
\end{figure}

\begin{figure}
\centering
\includegraphics[angle=0,width=8.8cm,height=7.6cm]{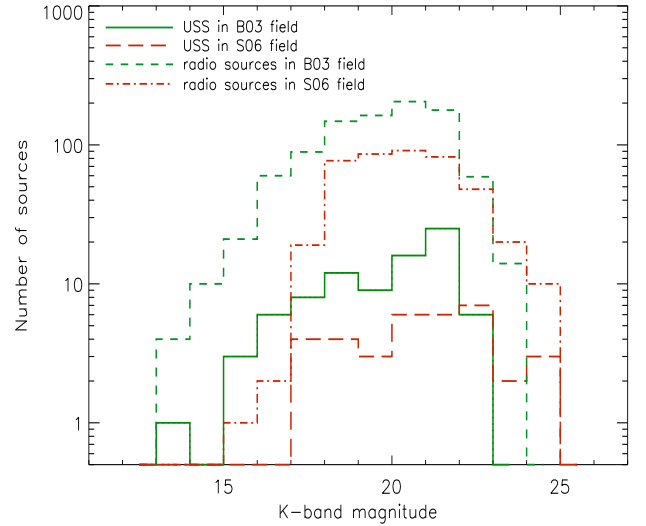}
\caption{Histograms of K-band magnitudes of the USS sources and of the full 1.4 GHz radio population in the B03 and the S06 field. 
Histograms of USS sources are shown by green solid lines and red long dashed lines for the B03 and the S06 fields, respectively. 
While green dashed and red dashed-dotted lines represent histograms for the full 1.4 GHz radio population in the B03 and the S06 fields, respectively.}
\label{fig:NearIRMagHist}
\end{figure}

\begin{figure}
\centering
\includegraphics[angle=0,width=8.8cm,height=8.0cm]{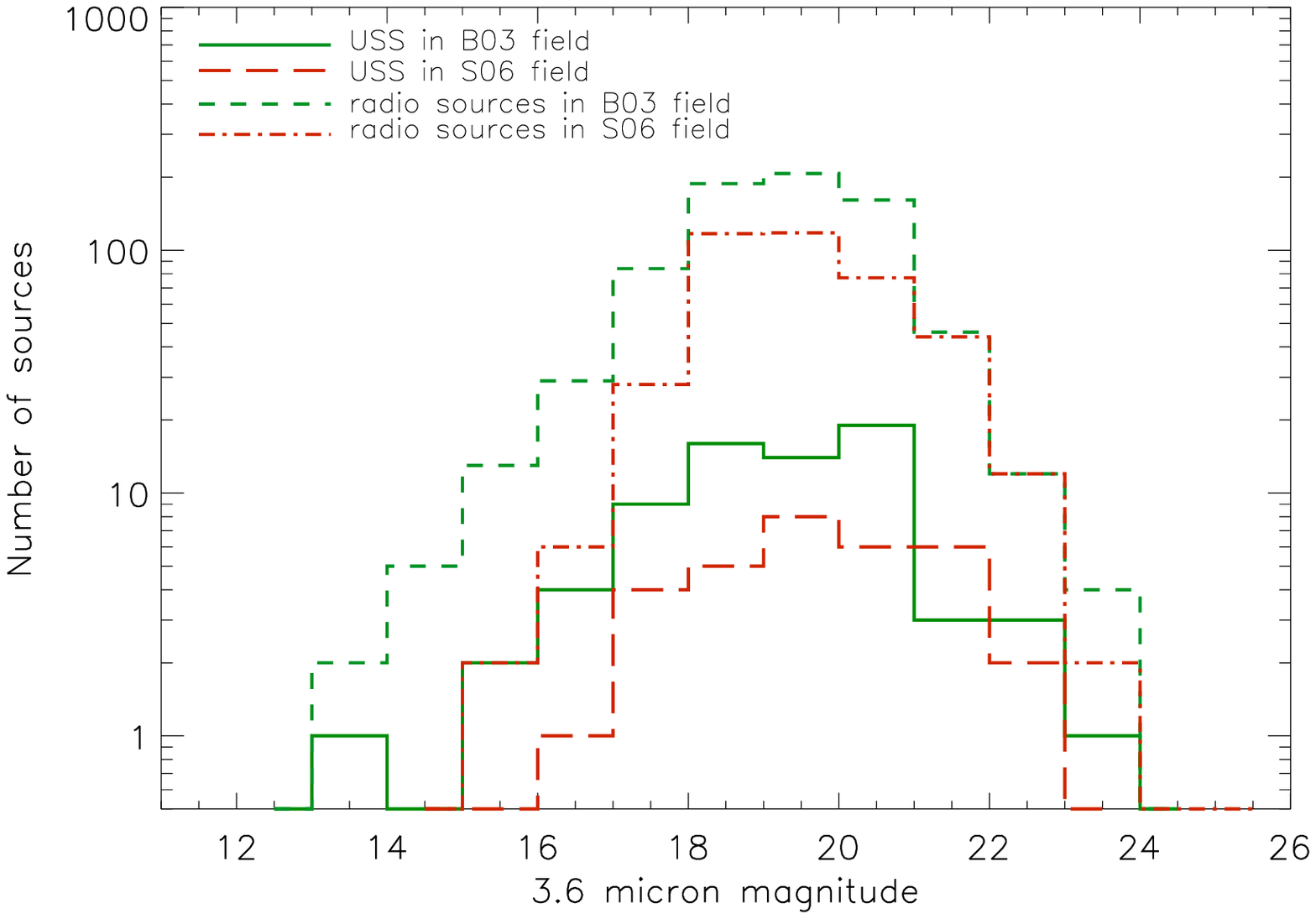}
\caption{Histograms of 3.6 $\mu$m magnitudes of the USS sources and of the full 1.4 GHz radio population in the B03 and S06 field. 
Histograms of USS sources are shown by green solid lines and red long dashed lines for the B03 and the S06 fields, respectively. 
While green dashed and red dashed-dotted lines represent histograms for full 1.4 GHz radio population in the B03 and the S06 fields, respectively.}
\label{fig:MidIRMagHist}
\end{figure}

Table~\ref{table:OpticalMags} lists the identification rates, medians and standard deviations of the optical, near-IR and mid-IR magnitude 
distributions for our USS sample sources as well as for the full radio population in the two subfields. 
The optical, near-IR and mid-IR counterparts of radio sources are found using likelihood ratio method 
and only counterparts with high reliability are considered as true counterparts ({\eg}\cite{Ciliegi05,Simpson06,McAlpine13}). 
We visually inspected near-IR/mid-IR images ({\eg}from VIDEO, UDS, SERVS, and SpUDS imaging) at the positions of all the USS sources 
and ensure that the counterparts found using the likelihood method are correct. 
The visual inspection at the positions of non-detections ({\ie}the USS sources without counterparts) shows that 
the majority of such sources remain undetected, except a few with either tentative faint counterparts at below 5$\sigma$ or 
lying close to a bright source. 
Also, the cross-matching of optical/near-IR/mid-IR sources with the 1.4 GHz radio sources shifted in random directions with 
random distances between 30 - 45 arcsec yields only $\sim$ 2$\%$ - 4$\%$ counterparts. 
This indicates that the false identification rate is limited only to a few percent level. 
\\
From Table~\ref{table:OpticalMags} it is evident that relatively less deep optical/near-IR/mid-IR surveys in the B03 field 
({\ie}Ks$_{\rm AB}$ $\leq$ 23.8) yields lower identification rate for USS sources ($\sim$ 74$\%$) compared to that for the full 
radio population ($\sim$ 89$\%$). 
While the use of deeper optical/near-IR/mid-IR data in the S06 field yields high and nearly similar identification rates 
({\ie}92$\%$) for both USS as well as for the full radio population. 
Previous studies have shown that the identification rates of bright USS sources with the optical/near-IR surveys limited 
to brighter magnitudes yield lower identification rates \citep{Wieringa91,Intema11}. 
However, deeper surveys result high identification rates for both the USS as well as non-USS sources \citep{DeBreuck02,Afonso11}.
Thus, our results on the optical/near-IR/mid-IR identification rates of our faint USS sources using existing deep surveys 
are consistent with previous findings.\\
Figure~\ref{fig:RMagHist}, Figure~\ref{fig:NearIRMagHist} and Figure~\ref{fig:MidIRMagHist}, respectively, show R band, K band 
and 3.6 $\mu$m magnitude distributions of our USS sources as well as of the full radio population, for both the subfields. 
We note that the optical/near-IR/mid-IR magnitude distributions of USS sources are flatter and have higher medians compared to the ones for 
the full radio population. This suggests that optical/near-IR/mid-IR counterparts of USS sources are systematically fainter compared to 
the ones for non-USS radio population. 
The two sample Kolmogorov$-$Smirnov (KS) test shows that the difference between the magnitude distributions of our USS sources 
and the full radio population increases 
at redder bands. The probability that null hypothesis is true {\ie}two sample have same distributions, decreases in red and 
IR bands ({\cf}Table~\ref{table:OpticalMags}). 
The two sample KS test on the comparison of the magnitude distributions of USS and non-USS radio sources give similar result. 
Thus, the comparison of optical/near-IR/mid-IR magnitude distributions of our USS sources and the full radio population is consistent with the 
interpretation that USS sources are relatively fainter and sample high-$z$ and/or dusty sources that have higher chances of 
being detected in the red/IR bands.
%
%
\begin{table*} 
\small
\caption{Average optical, near-IR and mid-IR magnitudes}
\scalebox{0.55}{\begin{tabular} {ccccccccccccccccccccc}
\hline
Band       & \multicolumn{20}{c}{Field} \\
                    & \multicolumn{10}{c}{B03} & \multicolumn{10}{c}{S06} \\ \hline  
                 & \multicolumn{3}{c}{USS radio sources}   & \multicolumn{3}{c}{All radio sources} &                &             & Depth   & Data & \multicolumn{3}{c}{USS radio sources} & \multicolumn{3}{c}{All radio sources}&                   &          & Depth  & Data \\ 
                 & identification       &   median & Std   & identification        & median& Std   & $\Delta$median & KS test     & at 5$\sigma$ & Ref. & identification    & median & Std   & identification       & median & Std   & $\Delta$median & KS test  & at 5$\sigma$    & Ref. \\ 
                 & rate                 &   Mag    &       & rate                  & Mag   &       &   Mag          & D (p-value) &   Mag        &     & rate              & Mag    &       & rate                  & Mag    &       &   Mag          & D (p-value) &  Mag  &    \\ \hline
Optical          &N$_{\rm USS}$ = 116   &          &       &N$_{\rm radio}$ = 1059 &       &       &                &             & A = 1.0      &  & N$_{\rm USS}$ = 39&        &       & N$_{\rm radio}$ = 512 &        &       &             &          & A = 0.8   &   \\
B                & 70 ($\sim$ 60.3$\%$) &   23.48  & 2.51  & 696 ($\sim$ 65.7$\%$) & 23.41 & 2.41  &  0.07          & 0.07 (0.85) & 26.5  & 1    & 36 ($\sim$ 92.3$\%$) & 24.13  & 2.54  & 481 ($\sim$ 93.9$\%$) & 23.90  & 2.26  &  0.23    & 0.13 (0.52)  & 28.4  & 2 \\
V                & 71 ($\sim$ 61.2$\%$) &   22.98  & 2.72  & 716 ($\sim$ 67.6$\%$) & 22.63 & 2.53  &  0.35          & 0.08 (0.83) & 26.2  & 1    & 36 ($\sim$ 92.3$\%$) & 23.38  & 2.56  & 483 ($\sim$ 94.3$\%$) & 22.98  & 2.24  &  0.40    & 0.15 (0.39)  & 27.8  & 2 \\
R                & 72 ($\sim$ 62.1$\%$) &   22.69  & 2.64  & 718 ($\sim$ 67.8$\%$) & 21.86 & 2.41  &  0.83          & 0.10 (0.54) & 25.9  & 1    & 36 ($\sim$ 92.3$\%$) & 23.92  & 2.81  & 493 ($\sim$ 96.3$\%$) & 23.01  & 2.48  &  0.91    & 0.20 (0.12)  & 27.7  & 2 \\
I                & 69 ($\sim$ 59.5$\%$) &   21.50  & 2.57  & 705 ($\sim$ 66.6$\%$) & 20.92 & 2.27  &  0.58          & 0.14 (0.14) & 25.0  & 1    &                      &        &       &                       &        &       &          &       &       &  \\
u$^{\star}$      & 73 ($\sim$ 62.9$\%$) &   23.80  & 2.19  & 780 ($\sim$ 73.7$\%$) & 24.00 & 2.19  & -0.20          & 0.09 (0.54) & 26.5  & 3    &                      &        &       &                       &        &       &          &       &       &  \\
g$^{\prime}$     & 85 ($\sim$ 73.3$\%$) &   23.48  & 2.43  & 879 ($\sim$ 83.0$\%$) & 23.44 & 2.41  &  0.04          & 0.09 (0.54) & 26.4  & 3    &                      &        &       &                       &        &       &          &       &       &  \\
r$^{\prime}$     & 86 ($\sim$ 74.1$\%$) &   22.98  & 2.51  & 899 ($\sim$ 84.9$\%$) & 22.75 & 2.48  &  0.23          & 0.10 (0.34) & 26.1  & 3    &                      &        &       &                       &        &       &          &       &       &  \\
i$^{\prime}$     & 86 ($\sim$ 74.1$\%$) &   22.43  & 2.52  & 918 ($\sim$ 87.1$\%$) & 21.96 & 2.45  &  0.47          & 0.08 (0.63) & 25.9  & 3    & 37 ($\sim$ 94.7$\%$) & 23.22  & 2.86  & 495 ($\sim$ 96.7$\%$) & 22.44  & 2.45  &  0.78    & 0.20 (0.11)   & 27.7  & 2 \\
z$^{\prime}$     & 83 ($\sim$ 71.5$\%$) &   21.86  & 2.44  & 897 ($\sim$ 85.1$\%$) & 21.36 & 2.33  &  0.50          & 0.10 (0.43) & 25.0  & 3    & 36 ($\sim$ 92.3$\%$) & 22.15  & 2.48  & 487 ($\sim$ 95.1$\%$) & 21.72  & 2.19  &  0.43    & 0.13 (0.53)   & 26.6  & 2 \\                     
%
near-IR          &N$_{\rm USS}$ = 116   &          &       &N$_{\rm radio}$ = 1059 &       &       &                &             & A = 1.0      & & N$_{\rm USS}$ = 38 &        &       &  N$_{\rm radio}$ = 459&        &       &          &        & A = 0.63   &  \\ 
Z                & 86 ($\sim$ 74.1$\%$) &   21.95  & 2.51  & 922 ($\sim$ 87.1$\%$) & 21.50 & 2.39  & 0.45           & 0.10 (0.39) & 25.7  & 4    &                      &        &       &                       &        &       &          &        &        &  \\
Y                & 82 ($\sim$ 70.7$\%$) &   21.33  & 2.36  & 890 ($\sim$ 84.0$\%$) & 20.97 & 2.20  & 0.36           & 0.11 (0.28) & 24.5  & 4    &                      &        &       &                       &        &       &          &        &        &  \\ 
J                & 85 ($\sim$ 73.3$\%$) &   20.98  & 2.27  & 927 ($\sim$ 87.5$\%$) & 20.69 & 2.10  & 0.29           & 0.11 (0.24) & 24.4  & 4    & 35 ($\sim$ 92.1$\%$) & 22.02  & 2.58  & 428 ($\sim$ 93.2$\%$) & 21.29  & 2.18  &  0.73    & 0.15 (0.50) & 24.9   & 5 \\ 
H                & 86 ($\sim$ 74.1$\%$) &   20.83  & 2.14  & 937 ($\sim$ 88.5$\%$) & 20.28 & 1.98  & 0.55           & 0.13 (0.14) & 24.1  & 4    & 35 ($\sim$ 92.1$\%$) & 21.34  & 2.40  & 430 ($\sim$ 93.7$\%$) & 20.82  & 2.06  &  0.52    & 0.15 (0.49) & 24.2   & 5 \\ 
K                & 86 ($\sim$ 74.1$\%$) &   20.40  & 2.01  & 951 ($\sim$ 89.8$\%$) & 19.87 & 1.87  & 0.53           & 0.11 (0.26) & 23.8  & 4    & 35 ($\sim$ 92.1$\%$) & 21.23  & 2.10  & 433 ($\sim$ 94.3$\%$) & 20.28  & 1.68  &  0.95    & 0.19 (0.22) & 24.6   & 5 \\ 
%
mid-IR           & N$_{\rm USS}$ = 95   &          &       & N$_{\rm radio}$ = 869 &       &       &                &            & A = 0.82     &  & N$_{\rm USS}$ = 36 &        &       & N$_{\rm radio}$ = 444 &        &       &          &         & A = 0.6   &  \\  
3.6 $\mu$m       & 72 ($\sim$ 75.8$\%$) &  19.57   & 1.72  & 751 ($\sim$ 86.4$\%$) & 19.27 & 1.44  &  0.30          & 0.09 (0.66) & 23.1  &  6   & 32 ($\sim$ 88.9$\%$) & 19.87  & 1.50  & 406 ($\sim$ 91.4$\%$) & 19.44  & 1.29  &  0.43    & 0.19 (0.26) & 24.0   & 7 \\ 
4.5 $\mu$m       & 70 ($\sim$ 73.7$\%$) &  19.46   & 1.62  & 746 ($\sim$ 85.8$\%$) & 19.39 & 1.28  &  0.07          & 0.11 (0.37) & 23.1  &  6   & 32 ($\sim$ 88.9$\%$) & 19.86  & 1.36  & 406 ($\sim$ 91.4$\%$) & 19.50  & 1.18  &  0.36    & 0.17 (0.41) & 24.0   & 7 \\ \hline
%
\end{tabular}}
\label{table:OpticalMags} 
\\
Notes -  B03 : \cite{Bondi03}; S06 : \cite{Simpson06}; Std : standard deviation; $\Delta$median = median (USS sources) - median (full radio population). \\
N$_{\rm USS}$ (N$_{\rm radio}$) represent total number of USS (radio) sources falling over the regions covered by the surveys at respective wavebands. 
`A' is the area in deg$^{2}$ of the overlapped region between the B03/S06 field and survey fields at respective wavebands.
Identification rate column gives number (percentage) of sources identified in the respective band.
Average magnitude errors in different bands are less than few percent (see references of respective surveys). 
Due to the unavailability of optical data, the optical identification rates in the S06 field do not include 5 USS sources of low signal-to-noise ratio ($<$5$\sigma$) at 1.4 GHz.
VIDEO K-band magnitudes are in K$_{\rm s}$ band. 
All the magnitudes are in AB system (UDS J, H, K magnitudes are converted from Vega to AB using conversion factors given in \cite{Hewett06}). \\ 
The two sample Kolmogorov$-$Smirnov (KS) test examines the hypothesis that two samples comes from same distribution. D = Sup x $|$S1(x) - S2(x)$|$ is the 
maximum difference between the cumulative distributions of two samples S1(x) and S2(x), respectively. The p-value is the probability that 
the null hypothesis, {\ie}two samples comes from same distribution, is correct.\\
References 1: VVDS data \citep{Ciliegi05}; 2: Subaru/Suprime-Cam data \citep{Simpson06}; 3: CFHTLS-D1 \citep{Ilbert06}; 
4: VIDEO survey \citep{Jarvis13}; 5: Ultra Deep Survey (UDS); 6: SERVS data \citep{Mauduit12}; 7: SpUDS data \citep{Dunlop07}.
\end{table*}


\section{Redshift distributions}
\begin{figure}
\centering
\includegraphics[angle=0,width=8.8cm,height=7.6cm]{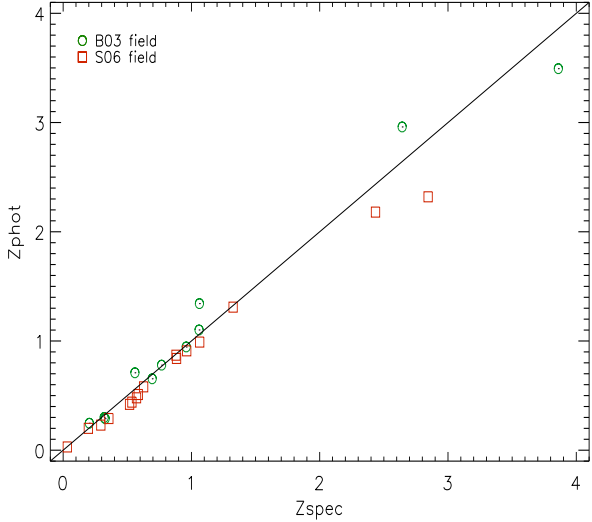}
\caption{Comparison between the spectroscopic and the photometric redshifts of the USS sources in the B03 field (green circle) 
and in the S03 field (red square). 
The diagonal line represents $z_{\rm spec}$ = $z_{\rm phot}$.}
\label{fig:ZspVsZph}
\end{figure}

\begin{figure}
\centering
\includegraphics[angle=0,width=8.8cm,height=7.6cm]{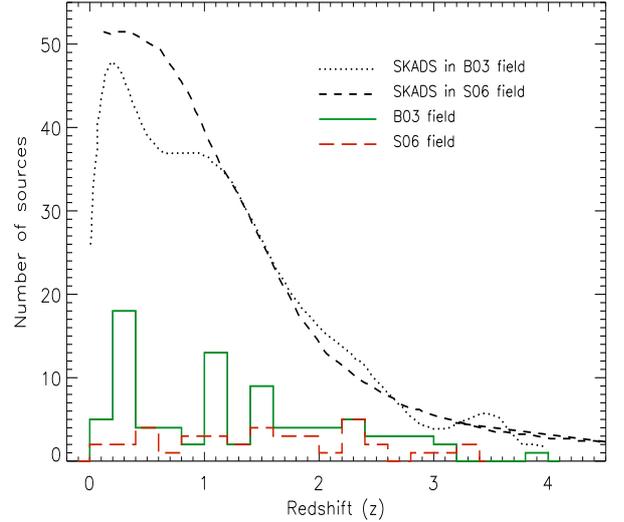}
\caption{Redshift distributions of our USS sources in the B03 field (in green solid lines) and in the S06 field (in red dashed lines). 
Redshift estimates are available for 86/116 and 39/44 USS sources in the B03 field and the S06 field, respectively. 
The redshift distributions of 1.4 GHz radio population predicted by SKA simulated skies (SKADS; \cite{Wilman08,Wilman10}) for the B03 
and the S06 fields are plotted with dotted and dashed curves, respectively. 
The flux limit S$_{\rm 1.4~GHz}$ $\sim$ 100 $\mu$Jy and sky area of 1.0 deg$^{-2}$ in the B03 field and 
0.8 deg$^{-2}$ in the S06 field, are used to obtain simulated radio population for the two subfields, respectively. 
The redshift distributions of simulated radio population in the B03 and the S06 fields are presented in 
\cite{McAlpine13} and \cite{Simpson12}, respectively. 
The uneven variations seen in the SKADS simulated redshift distribution in the B03 field can be attributed to the clustering of radio sources 
manifested as cosmic variance in this relatively small field.}
\label{fig:RedshiftHist}
\end{figure}

To obtain redshifts of our USS sample sources, we use the spectroscopic and photometric measurements available in the literature. \\
{\bf The B03 field} : There has been more than one attempt to estimate photometric redshifts of the 1.4 GHz radio sources in the B03 field 
({\eg}\cite{Ciliegi05,Bardelli09,McAlpine13}).
Using deep 10$-$bands photometric data ({\ie}five bands near-IR VIDEO data combined with five bands CFHTLS-D1 optical data) \cite{McAlpine13} present 
most accurate photometric redshift estimates of 1.4 GHz radio sources. 
The photometric redshifts were determined using the code 
Le Phare\footnote{http://www.cfht.hawaii.edu/ arnouts/LEPHARE/lephare.html} \citep{Ilbert06} that uses a trial of 
fitting the photometric bands with a set of input Spectral Energy Distribution (SED) templates. 
The accuracy of the photometric redshifts was assessed by comparing with secure spectroscopic redshifts obtained with the VIMOS
VLT deep survey (VVDS; \cite{LeFevre05}). 
Approximately 3.8 per cent of the sources are catastrophic outliers, defined as cases with $\Delta$z / (1+z$_{\rm s}$) $>$ 0.15, 
where $\Delta$z = $|$z$_{\rm p}$~$-$ z$_{\rm s}$$|$. 
The details of the procedure used to derive these photometric redshifts are given in \cite{Jarvis13}. \\
Using photometric redshift estimates from \cite{McAlpine13}, we find that 86/116 USS sources in the B03 field 
have photometric redshifts. 
Nearly 0.64 deg$^{2}$ of the B03 field is also covered by the VVDS which is a magnitude limited spectroscopic 
redshift survey conducted by the VIMOS multi-slit spectrograph at the ESO-VLT \citep{LeFevre13}. 
Using the latest VVDS catalog\footnote{http://cesam.lam.fr/vvds}, we find that 
only 11 USS sources have spectroscopic redshifts, and all these sources also have photo-$z$ estimates from \cite{McAlpine13}.
There are 30/116 ($\sim$ 25.8$\%$) USS sources without redshift estimates and these may potentially be high redshift candidates that are too
faint to be detected in existing optical, IR surveys. \\
{\bf The S06 field} : \cite{Simpson12} present spectroscopic and 11$-$band (u$^{\star}$, B, V, R, i$^{\prime}$, z$^{\prime}$, J, H, K plus IRAC bands 1 and 2) 
photometric redshifts for 505/512 1.4 GHz radio sources. 
The spectroscopic redshift measurements are obtained using the Visible Multi-Object Spectrograph (VIMOS) on the VLT and 
also include measurements from different spectroscopic campaigns in the SXDF field 
({\eg}\cite{Geach07,Smail08,VanBreukelen09,Banerji11,chuter11}, Pearce et al. (in preparation), Akiyama et al. (in preparation)).
Spectroscopic redshifts are available for 267/505 radio sources, while rest of the radio sources have photometric redshift estimates.
The photometric redshifts were estimated using the code EAZY \citep{Brammer08} after correcting the observed
photometry for Galactic extinction of A$_{\rm V}$ = 0.070 \citep{Schlegel98} with the Milky Way extinction
law of \cite{Pei92}. 
Using \cite{Simpson12} redshifts measurements we find that spectroscopic redshifts are available for 16/44 USS sources, 
while 23/44 USS sources have photometric redshifts.  
We compare the spectroscopic redshifts ($z_{\rm spec}$) and the photometric redshifts ($z_{\rm phot}$) for 
all those USS sources that have both types of redshift estimates.  
Figure~\ref{fig:ZspVsZph} shows the comparison of $z_{\rm spec}$ and $z_{\rm phot}$ and it is clear that the $z_{\rm phot}$ estimates 
are fairly consistent with the $z_{\rm spec}$ measurements at $z \leq 1.5$. They are less accurate at higher redshifts. 
We do not see any catastrophic outliers in the comparison of spectroscopic redshifts ($z_{\rm spec}$) and photometric redshifts ($z_{\rm phot}$), 
although this comparison is limited only to a small fraction of our USS sources. \\
Figure~\ref{fig:RedshiftHist} shows the redshift distributions of our USS sources in the two subfields. 
We use spectroscopic redshifts whenever available, otherwise photometric redshifts are used.
The USS redshift distribution in the B03 field spans from 0.096 to 3.86 with mean ($z_{\rm mean}$) $\sim$ 1.31 and 
median ($z_{\rm median}$) $\sim$ 1.18.
It is evident that substantially large fraction (53/86 $\sim$ 61.5$/\%$) of USS sources in the B03 field, are lying at $z \geq 1.0$. 
The USS redshift distribution in the S06 field is flatter and spans from 0.033 to 3.34 with $z_{\rm mean}$ $\sim$ 1.54 
and $z_{\rm median}$ $\sim$ 1.57.
We note that 27/44 $\simeq$ 61.4$\%$ of USS sources in the S06 field are at redshifts ($z$) $\geq$ 1.0.
The lower median redshift of the USS sample in the B03 field can be attributed to the fact that there are no redshift estimates 
for a significantly large fraction (30/116 $\sim$ 25.8$\%$) of USS sources in this field. 
The USS sources without redshifts remained undetected in the existing optical, 
IR surveys and may possibly be faint sources at higher redshifts. We discuss the possible nature of these USS sources in the Section 5.2. \\ 
The USS redshift distribution in the B03 field also shows peaks at $z$ $\sim$ 0.3, $z$ $\sim$ 1.2 and at $z$ $\sim$ 1.5. 
It is to be noted that the redshift distribution of near-IR identified radio sources also exhibits peak at $z$ $\sim$ 0.2 $-$ 0.4 and $z$ $\sim$ 1.0 $-$ 1.2 
\citep{McAlpine13}. The redshift peak at $z$ $\sim$ 0.2 $-$ 0.4 can plausibly be due to large-scale structure within this relatively small field 
{\ie}there are six known X-ray clusters at $z$ $\simeq$ 0.262, 0.266, 0.293, 0.301, 0.307 and 0.345 
\citep{Pacaud07,Adami11} present in this field, which is at least partially responsible for an increase in the sources in this redshift range. 
We surmise that the redshift peaks at $z$ $\sim$ 1.2 and 1.5 may also be due to the presence of clusters at these redshifts, although we caution 
that the majority of redshift estimates are based on photometry.
\\
In order to examine whether our USS sample indeed selects high-$z$ sources, we compare median redshift of our USS sources 
with that of the non-USS sources.   
The 325 MHz $-$ 1.4 GHz cross-matched catalog yields 227 and 152 non-USS sources (${\alpha}_{\rm 325~MHz}^{\rm 1.4~GHz}$ $<$ 1.0) 
in the B03 and the S06 field, respectively. 
We find that only 192/227 ($\sim$ 84.6$\%$) and 135/152 ($\sim$ 88.8$\%$) do have redshift estimates with the median redshift values $\sim$ 0.99 and 
$\sim$ 0.96, in the B03 and the S06 field, respectively. 
It is evident that on average the USS sources ($z_{\rm median}$ $\sim$ 1.18 in the B03 field and $z_{\rm median}$ $\sim$ 1.57 in the S06 field) 
are at higher redshifts than the non-USS radio sources.
To check, if within the USS sample, the radio sources with relatively steeper spectral index are at relatively higher redshifts, 
we make two subsamples of USS sources {\ie}one consists of sources with ${\alpha}_{\rm 325~MHz}^{\rm 1.4~GHz}$ $\leq$ -1.3, 
and the other USS subsample consists of sources with -1.3 $<$ ${\alpha}_{\rm 325~MHz}^{\rm 1.4~GHz}$ $\leq$ -1.0. 
We find that, in the B03 field, among the 86/116 sources with available redshifts only 22/86 USS sources 
have ${\alpha}_{\rm 325~MHz}^{\rm 1.4~GHz}$ $\leq$ - 1.3 and yield median redshift of $\sim$ 1.72, while 64/86 USS sources 
with -1.3 $<$ ${\alpha}_{\rm 325~MHz}^{\rm 1.4~GHz}$ $\leq$ -1.0 have median redshift of $\sim$ 1.08. 
In the S06 field, among the 39/44 USS sources with available redshifts only 5 USS sources have 
${\alpha}_{\rm 325~MHz}^{\rm 1.4~GHz}$ $\leq$ -1.3 with the median redshift $\sim$ 1.32, while 34 USS sources 
with -1.3 $<$ ${\alpha}_{\rm 325~MHz}^{\rm 1.4~GHz}$ $\leq$ -1.0 have median redshift $\sim$ 1.57. 
It is to be noted that, in the S06 field, the number of USS sources with ${\alpha}_{\rm 325~MHz}^{\rm 1.4~GHz}$ $\leq$ -1.3 are not sufficient
to make a robust statistical comparison. 
Therefore, based on the USS sources in the B03 field, we find that, on average, sources with steeper radio spectral index tend to have higher redshift. 
This result is consistent with the $z-{\alpha}$ correlation \citep{Ker12}. 
\\
We also compare the redshift distribution of our USS sources with the one for 
the radio population derived by using the SKADS Simulated Skies (S$^{3}$) simulations \citep{Wilman08,Wilman10} ({\cf}Figure~\ref{fig:RedshiftHist}). 
The S$^{3}$ simulation uses a model which includes different radio populations 
{\ie}star$-$forming galaxies, radio$-$quiet AGNs, radio$-$loud AGNs (FR$-$I and FR$-$II radio galaxies). 
The S$^{3}$$-$simulations$\footnote{http://s-cubed.physics.ox.ac.uk/}$ do not cover 325 MHz frequency which is the base frequency of 
our USS sample, and therefore we use 1.4 GHz frequency to obtain the redshift distribution of the simulated radio population. 
Figure~\ref{fig:RedshiftHist} shows that the redshift distributions of simulated 1.4 GHz radio population peak 
at low redshift with a sharp decline over $z$ $\sim$ 1 to 3 and a nearly flat tail at $z$ $>$ 3.0. 
In contrast to the simulated radio population, the redshift distributions of USS sources in the two subfields are nearly flat, 
except for the two peaks seen in the B03 field that are possibly attributed to the presence of galaxy clusters in this field. 
The difference between the redshift distributions of USS sources and the simulated radio population is maximum at low redshift, while it 
decreases at higher redshifts, particularly at $z$ $\geq$ 2.0. 
This suggests that the USS technique preferentially selects high-$z$ sources, while removing a large fraction of low-$z$ sources. 
At sub-mJy flux densities, the radio population is known to be dominated by star$-$forming galaxies and low-power AGNs with 
increasing contribution by AGNs at higher redshifts \citep{Wilman08,Wilman10}.  
Thus, in our faint USS sample, the high-$z$ radio sources are likely to be dominated by relatively low-power AGNs such as FR$-$I radio galaxies. 
However, powerful FR$-$II radio galaxies at even higher redshifts can also be present in our USS sample.

\begin{table*} 
\caption{The USS sample parameters}
\begin{minipage}{140mm}
\begin{tabular} {ccccccccccc}
\hline
                   & \multicolumn{5}{c}{B03 field} & \multicolumn{5}{c}{S06 field} \\ \hline
Parameter            & No. of sources & Min   & Max   & Median & Std   & No. of sources & Min  & Max   & Median & Std\\ \hline
S$_{\rm 325~MHz}$ (mJy)    &   116    & 0.484 & 108.8 &  1.76  & 16.2 &  44            & 0.51 & 367.9 & 1.96  & 68.1  \\
S$_{\rm 1.4~GHz}$ (mJy)    &   116    & 0.070 & 18.96 &  0.27  & 3.30 &  44            & 0.076 & 80.3  & 0.36 & 13.9 \\
logL$_{\rm 1.4~GHz}$ (W~Hz$^{-1}$) & 86/116        & 21.46 & 26.07 & 25.50  & 1.10 &  39/44       & 21.52 & 27.43 & 24.86  & 1.13 \\
logL$_{\rm 325~MHz}$ (W~Hz$^{-1}$) & 86/116        & 22.23 & 26.71 & 25.29  & 1.12 &  39/44       & 22.17 & 28.13 & 25.57  & 1.16 \\  
Redshift              & 86/116 (11)   & 0.097 & 3.86  & 1.18   & 0.91 &  39/44 (16)  & 0.033 & 3.34  & 1.57   & 0.86  \\ \hline
\end{tabular} 
\label{tab:USSParam} 
\vspace {0.1cm} \\
Notes -  B03 : \cite{Bondi03}; S06 : \cite{Simpson06}. Number of sources with spectroscopic redshifts are mentioned inside brackets.
\end{minipage}
\end{table*}

\section{Color$-$color diagnostics}
In order to understand the nature of USS sources in our sample we investigate the mid-IR colors and the flux ratios of radio to mid-IR.
\subsection{Mid-IR colors}
\begin{figure*}
\centering
\includegraphics[angle=0,width=8.8cm,height=7.6cm]{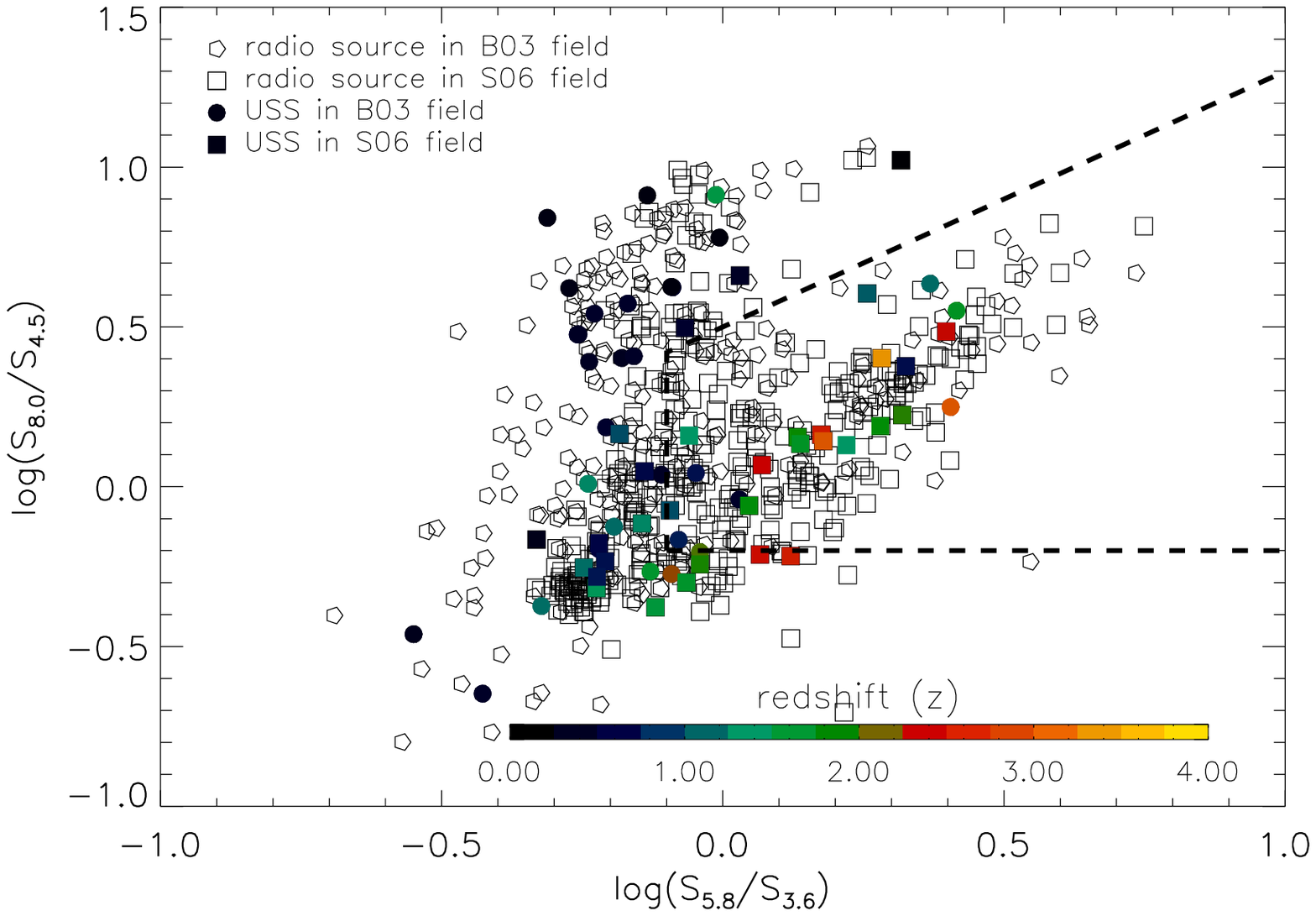}
\includegraphics[angle=0,width=8.8cm,height=7.6cm]{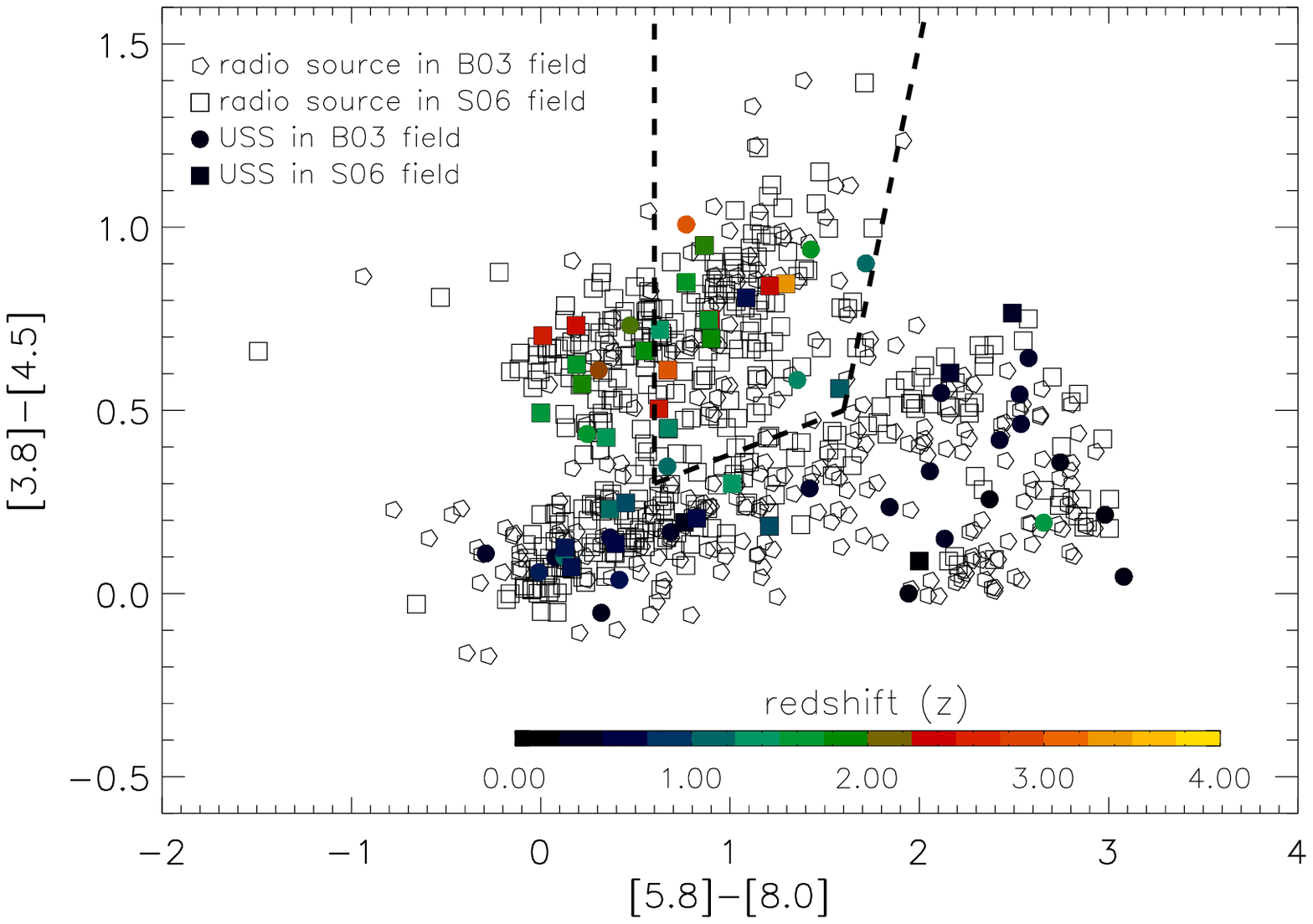}
\caption{Mid-IR color-color diagnostic plots for our USS sources in both the B03 and S06 fields. 
Left and right panels show mid-IR color-color plots based on \cite{Lacy04} and \cite{Stern05} criteria, respectively. 
Filled and open symbols represent USS sources and 1.4 GHz radio population, respectively. 
USS sources of different redshifts are shown with different colors. The regions bounded by dashed lines denote AGN selection wedge.  
\cite{Lacy04} defined the AGN selection wedge as : 
(log(S$_{\rm 5.8}$/S$_{\rm 3.6}$) $>$ -0.1) $\wedge$ (log(S$_{\rm 8.0}$/S$_{\rm 4.5}$) $>$ -0.8) $\wedge$ 
(log(S$_{\rm 8.0}$/S$_{\rm 4.5}$) $\leq$ 0.8 log(S$_{\rm 5.8}$/S$_{\rm 3.6}$) + 0.5); where $\wedge$ is `AND' operator. 
While, the AGN selection wedge proposed by \cite{Stern05} is defined as : 
([5.8] - [8.0] $>$ 0.6) $\wedge$ ([3.6] - [4.5] $>$ 0.2([5.8] - [8.0]) + 0.18) $\wedge$ ([3.6] - [4.5] $>$ 2.5 ([5.8] - [8.0]) - 3.5); where 
IRAC magnitudes are in the Vega system. 
We converted IRAC AB magnitudes to Vega magnitudes (m$_{\rm AB}$ = m$_{\rm Vega}$ + conv) 
using conversion factors 2.78, 3.26, 3.75 and 4.38 for 3.6 $\mu$m, 4.5 $\mu$m, 5.8 $\mu$m and 8.0 $\mu$m bands, respectively 
(see IRAC Data HandBook 3.0 2006).}
\label{fig:MidIRcolor}
\end{figure*}

Mid-IR Spectral Energy Distributions (SEDs) of AGN are generally characterized by a power law and differ from star$-$forming galaxies \citep{Alonso-Herrero06,Donley07}. 
Therefore, mid-IR colors are useful in identifying the presence of AGN-heated dust in the SEDs of galaxies. 
We investigate the nature of our USS sample sources using mid-IR color diagnostics proposed by \cite{Lacy04} and \cite{Stern05}.
We note that only 32/116 (27.6$\%$) USS sources in the B03 field and 32/44 (72.7$\%$) USS sources in the S06 field 
have detections in all four IRAC bands (3.6, 4.5, 5.8 and 8.0 $\mu$m) from the SWIRE and 
the SpUDS data, respectively. 
Thus, the mid-IR color-color diagnostic is limited only to a fraction of our USS sample sources. 
The higher fraction of USS sources detected in the S06 field may be attributed to the deeper SpUDS data 
(5$\sigma$ depth at 3.6~$\mu$m $\sim$ 0.9~$\mu$Jy) compared to the SWIRE (5$\sigma$ depth at 3.6~$\mu$m $\sim$ 3.7~$\mu$Jy). \\
Figure~\ref{fig:MidIRcolor} shows mid-IR color-color diagnostic plots for our USS sample sources as well as 
for the radio population in the two subfields. 
The MIR color-color diagnostic plots based on \cite{Lacy04} and \cite{Stern05} criteria show that our USS sources exhibit 
wide range of mid-IR colors with large fraction of USS sources falling in the AGN selection wedge. 
However, in the B03 field, nearly half of the USS sample sources reside outside the AGN selection wedge.
Notably, most of the USS sources lying outside of the AGN wedge selection are of low redshifts ($z {\leq} 0.5$). 
Therefore, low-$z$ USS sources of our sample, particularly in the B03 field, are likely to be contaminated by star$-$forming galaxies 
or composite galaxies in which IR emission is dominated by star formation. 
We note that our mid-IR color diagnostic, in the B03 field, is based on the relatively shallow SWIRE data which is 
expected to detect relatively bright sources. 
The USS sources at higher redshifts ($z > 0.5$), in both the subfields, preferentially fall either inside or close to the AGN selection wedge.  
Thus, MIR color-color diagnostics are consistent with a large fraction of our USS sample sources at relatively higher redshifts ($z > 0.5$) 
being mainly AGN. 
However, due to non-detection of a substantial fraction of USS sources in all four IRAC bands, 
we cannot obtain the exact fraction of AGN dominated USS sources in our sample. 
Furthermore, we caution that the mid-IR color-color diagnostic plots are known to be contaminated {\ie}AGN may fall in non-AGN regions and vice-versa 
(see, \cite{Donley08,Barmby08,Donley12}).
In fact, the samples of radio$-$loud AGN are known to exhibit wide variety of IR colors with dichotomy displayed in mid-IR-radio plane for low and 
high excitation radio galaxies (see \cite{Gurkan14}).
Also, there are suggestions that radio selected AGNs may have different accretion mode {\ie}radiatively inefficient (`radio mode'), and may not 
strictly follow the mid-IR color selection criteria \citep{Croton06,Hardcastle07,Tasse08,Griffith10}. 
\\
\cite{Simpson12} present optical spectra of 267/512 radio sources detected at 1.4 GHz in the S06 field. 
Our USS sources are a sub-sample of the 1.4 GHz radio sources and we find that optical spectra are available for 15 USS sources.
Spectral classifications based on observed emission and/or absorption line properties shows that 
five USS sources are Narrow Line AGN (NLAGN), five USS sources are Star Burst (SB), 
three and one USS sources are, respectively, strong and weak line emitter with uncertain classification, and one source is classified 
as a absorption line galaxy.    
We note that the USS sources classified as starburst galaxies are preferentially at lower redshifts ($z < 0.5$), while NLAGNs are at higher redshifts 
($z > 0.5$), which is consistent with the findings of our mid-IR color-color diagnostic.

\subsection{Flux ratios of radio to mid-IR}
\begin{figure}
\centering
\includegraphics[angle=0,width=9.2cm,height=8.2cm]{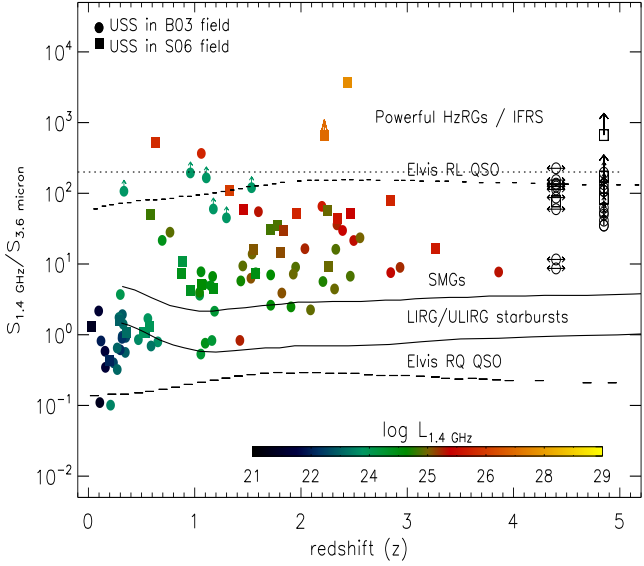}
\caption{Ratio of 1.4~GHz radio flux density to 3.6~$\mu$m flux (S$_{\rm 1.4~GHz}$/S$_{\rm 3.6~{\mu}m}$) versus 
redshift plot for our USS sources. USS sources of different radio luminosities are represented with different colors. 
USS sources without redshift estimates are shown at the rightmost position. USS sources with only lower limits on the flux ratios 
(S$_{\rm 1.4~GHz}$/S$_{\rm 3.6~{\mu}m}$) {\ie}without 3.6 $\mu$m detections, are shown by upward arrows.  
The topmost area above the dotted line represents the range of flux ratios for the powerful H$z$RGs and IFRSs. 
Tracks indicating the regions for the different class of sources are taken from \cite{Norris11}. 
The solid lines represent the loci of LIRGs and ULIRGs using \cite{Rieke09} SED templates. The dashed (long dashed) line indicates the loci of 
radio$-$loud (radio$-$quiet) QSOs from \cite{Elvis94}. We caution that dust extinction can cause any of these tracks to rise steeply at high redshift, 
where the observed 3.6 $\mu$m is emitted in visible wavelengths at the rest frame. 
}
\label{fig:FluxRatioVsRedshift}
\end{figure}

The ratio of 1.4 GHz flux density to 3.6 $\mu$m flux (S$_{\rm 1.4~GHz}$/S$_{\rm 3.6~{\mu}m}$) versus redshift plot can be used as a diagnostic 
to differentiate sources of different classes {\ie}star$-$forming galaxies, radio$-$quiet AGN, H$z$RGs (see, \cite{Norris11}). 
In general, H$z$RGs and radio$-$loud AGNs exhibit high ratio of 1.4 GHz 
flux density to 3.6 $\mu$m flux (S$_{\rm 1.4~GHz}$/S$_{\rm 3.6~{\mu}m}$), while radio$-$quiet and star$-$forming galaxies are characterized by a low ratio.
Figure~\ref{fig:FluxRatioVsRedshift} shows the ratio of 1.4 GHz flux density to 3.6 $\mu$m flux (S$_{\rm 1.4~GHz}$/S$_{\rm 3.6~{\mu}m}$) 
versus redshift plot for our USS sample sources. 
We note that the radio to mid-IR flux ratio diagnostic is limited only to those USS sources which are covered by SERVS ({\eg}95/116 USS in the B03 field) 
and SpUDS ({\eg}36/44 USS in the S06 field) survey regions ({\cf}Figure~\ref{fig:footprints}). 
In our sample, 72/95 USS sources in the B03 field and 32/36 USS sources in the S06 field, do have 3.6 $\mu$m counterpart 
({\cf}Table~\ref{table:OpticalMags}). 
While, for USS sources without 3.6 $\mu$m detections ({\ie}16 sources in the B03 field and four sources in the S06 field), we put a lower limit 
on the flux ratio S$_{\rm 1.4~GHz}$/S$_{\rm 3.6~{\mu}m}$ using 3.6 $\mu$m survey flux limits ({\ie}2.0 $\mu$Jy for the SERVS data and 0.9 $\mu$Jy for the SpUDS data). 
We note that in the B03 field there are 7 USS sources with extended radio sizes for which 3.6 $\mu$m counterparts are unavailable due to ambiguity 
caused by the existence of more than one IRAC source detected within their radio sizes. 
These sources are not included in the flux ratio diagnostic plot. \\
From Figure~\ref{fig:FluxRatioVsRedshift}, it is evident that our USS sample sources in both the subfields are distributed over a wide range 
of flux ratios (S$_{\rm 1.4~GHz}$/S$_{\rm 3.6~{\mu}m}$ $\sim$ 0.1 - 1000) and redshifts ($z~{\sim} 0.1 - 3.8$). 
The flux diagnostic plot also shows tracks indicating regions of different class of sources as proposed by \cite{Norris11}. 
From the flux diagnostic plot, it is clear that our USS sample contains sources of various classes.     
At low redshifts ($z~{\leq}~0.5$), most of our USS sources tend to exhibit low ratio of radio to mid-IR 
({\ie}S$_{\rm 1.4~GHz}$/S$_{\rm 3.6~{\mu}m}$ $\leq$ 1.0) and low radio luminosities (L$_{\rm 1.4~GHz}$ $<$ 10$^{24}$ W Hz$^{-1}$) 
({\cf}Figure~\ref{fig:FluxRatioVsRedshift}), 
similar to star$-$forming galaxies and radio$-$quiet AGNs. 
This is consistent with the mid-IR color-color diagnostic in which low-$z$ USS sources 
tend to lie outside the AGN selection wedge. 
The presence of low-$z$ star$-$forming galaxies in a faint USS sample is not unexpected, as the dominant non-thermal radio 
emission at low-frequencies can give rise spectral index as steep as -1.0 \citep{Heesen09,Basu12}. \\           
In the flux ratio diagnostic plot, a small fraction of USS sources (10/88 $\sim$ 11$\%$ sources in the B03 field 
and 2/36 $\sim$ 5.5$\%$ sources in the S06 field) are found to be distributed between the flux ratio tracks of Luminous IR Galaxies (LIRGs) 
and Ultra Luminous IR Galaxies (ULIRGs) starbursts 
({\cf}Figure~\ref{fig:FluxRatioVsRedshift}). 
The typical radio luminosities of these USS sources are L$_{\rm 1.4~GHz}$ $\sim$ 10$^{23}$ $-$ 10$^{25}$ W Hz$^{-1}$. 
The relatively high radio luminosities and the steep radio spectral index can be considered as the indication of the presence of AGN. 
In fact, some of LIRGs/ULIRGs are known to host AGNs \citep{Risaliti10,Lee12} which are detected in deep radio observations \citep{Fiolet09,Leroy11}. 
Therefore, a fraction of our USS sources are likely to be obscured AGNs hosted in LIRGs/ULIRGs. 
Furthermore, there is a substantially large fraction of our USS sample sources 
(33/88 $\sim$ 38$\%$ in the B03 field and 21/36 $\sim$ 58$\%$ in the S06 field) with the locations in the flux 
diagnostic plot similar to the ones observed for Sub-Millimeter Galaxies (SMGs) in the representative sample of \cite{Norris11}.  
These USS sources are distributed over redshift $\sim$ 0.5 to 3.8 with flux ratios (S$_{\rm 1.4~GHz}$/S$_{\rm 3.6~{\mu}m}$) $\sim$ 4 to 100 and 
radio luminosities L$_{\rm 1.4~GHz}$ $>$ 10$^{24}$ W Hz$^{-1}$. 
The high radio luminosities and steep spectral index can be indicative of the presence of possible radio$-$loud AGN.  
Indeed, a few ULIRGs, SMGs at $z$ $\sim$ 2.0 are known to host radio$-$loud AGNs often characterized with ultra 
steep radio spectrum ({\eg}\cite{Sajina07,Polletta08,Martinez09}). 
The heavily obscured radio$-$loud AGNs are, in general, faint USS sources ({\ie}S$_{\rm 1.4~GHz}$ $\sim$ 0.5 $-$ 2.0 mJy, 
${\alpha}_{\rm 610 MHz}^{\rm 1.4~GHz}$ $\leq$ -1.0 ({\eg}\cite{Sajina07,Ibar10})), similar to the ones present in our USS sample.
These sources are believed to be heavily obscured AGNs, observed in the transition stage 
after the birth of the radio source, but before feedback effects dispel the interstellar medium and halt the starburst activity.
Few of the local ULIRGs ({\eg}F00183-7111) are known to show a compact radio core-jet AGN with radio luminosity typical of powerful radio galaxies 
({\eg}\cite{Norris12}). 
Thus, radio to mid-IR flux ratio diagnostic implies that a substantially large fraction  
(more than one third in the B03 field and two third in the S06 field) of our faint USS sample sources are likely to be 
relatively weaker radio$-$loud AGNs (L$_{\rm 1.4 GHz}$ $\sim$ 10$^{24}$ $-$ 10$^{26}$ W Hz$^{-1}$) hosted in obscured environments of ULIRGs and SMGs.
Some of the USS sources in the S06 field classified as NLAGN \citep{Simpson12} 
have flux ratio of radio to mid-IR similar to ULIRGs/SMGs and therefore these 
sources can be type 2 AGN hosted in dusty obscured environments (see \cite{Martnez05,Donley05,Martinez09}).   
\\      
There is a fraction of USS sources (10/88 $\sim$ 11$\%$ in the B03 field) that are detected at 3.6 $\mu$m but remain undetected at near-IR and optical 
and therefore do not have redshift estimates. 
In the flux diagnostic plot, these sources are shown at rightmost location with the horizontal two-sided arrows. 
Most of these USS sources have high flux ratios of radio to mid-IR (S$_{\rm 1.4~GHz}$/S$_{\rm 3.6~{\mu}m}$ $>$ 50) and 
are candidate H$z$RG.
Furthermore, there is a significant fraction of USS sources (16/88 $\sim$ 18$\%$ in the B03 field and 3/36 $\sim$ 8.3$\%$ in the S06 field) 
that do not have 3.6 $\mu$m detections, 
and therefore only lower limits on the flux ratios S$_{\rm 1.4~GHz}$/S$_{\rm 3.6~{\mu}m}$ are assigned. 
Most of these USS sources do not have optical and near-IR detections too, and therefore, no redshift estimates are available.
These USS sources are shown at the rightmost location with upward arrows in the flux diagnostic plot and have radio to mid-IR flux ratio limits 
(S$_{\rm 1.4~GHz}$/S$_{\rm 3.6~{\mu}m}$) $>$ 50. 
We note that seven USS sources with extended radio sizes lack reliable 3.6 $\mu$m counterparts and would have much high flux ratio limits 
({\ie}S$_{\rm 1.4~GHz}$/S$_{\rm 3.6~{\mu}m}$ $>$ 200), if their 3.6 $\mu$m counterparts are undetected. 
Recent studies have reported the existence of radio sources with faint or no IR counterparts, termed as Infrared Faint Radio Sources (IFRS) 
(see \cite{Norris11}), which show high flux ratio of 1.4 GHz to 3.6 $\mu$m {\ie}S$_{\rm 1.4~GHz}$/S$_{\rm 3.6~{\mu}m}$ $>$ 50, 
and many IFRSs are known to exhibit ultra steep radio spectrum ({\eg}\cite{Middelberg11}).  
Follow-up studies of IFRS sources suggest that majority of these sources are obscured high-$z$ radio$-$loud AGNs, 
possibly suffering from significant dust extinction \citep{Norris07,Norris11,Middelberg08,Huynh10,Collier14}. 
Thus, our flux ratio diagnostic infers that we have a significant fraction of USS sample sources 
(26/88 $\sim$ 29.5$\%$ in the B03 field and 4/36 $\sim$ 11$\%$ in the S06 field) as IFRSs, which in turn are also potential H$z$RG candidates.
\\
Furthermore, we note that the flux ratio of 1.4 GHz to 3.6 $\mu$m (S$_{\rm 1.4~GHz}$/S$_{\rm 3.6~{\mu}m}$) versus redshift ($z$) diagnostic plot 
suggests that a high cut-off in S$_{\rm 1.4~GHz}$/S$_{\rm 3.6~{\mu}m}$ can be used to select high-$z$ sources. 
For example, contamination by low-$z$ star$-$forming galaxies in our USS sample can be completely removed if 
we take S$_{\rm 1.4~GHz}$/S$_{\rm 3.6~{\mu}m}$ $>$ 10. 
Using S$_{\rm 1.4~GHz}$/S$_{\rm 3.6~{\mu}m}$ $>$ 10 yields only high-$z$ sources ($z$ $\geq$ 1) and few radio-strong AGN at lower redshifts. 
This is consistent with the fact that IFRSs, candidate H$z$RG, are characterized with high flux ratio 
of 1.4 GHz to 3.6 $\mu$m (S$_{\rm 1.4~GHz}$/S$_{\rm 3.6~{\mu}m}$ $>$ 50; {\eg}\cite{Norris11,Collier14}).

\section{Radio luminosities of USS sources}
\begin{figure}
\centering
\includegraphics[angle=0,width=8.8cm,height=7.8cm]{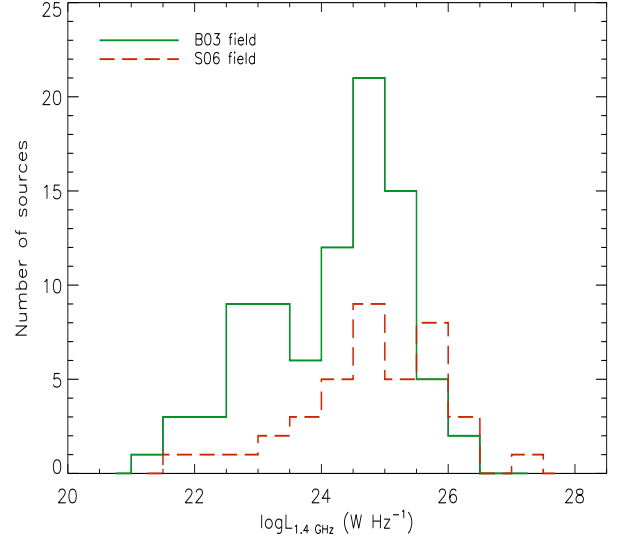}
\caption{Histograms of 1.4 GHz radio luminosities of our USS sample sources in the B03 (green solid lines) and in the S06 (red dashed lines) fields.}
\label{fig:Lradio}
\end{figure}

\begin{figure}
\centering
\includegraphics[angle=0,width=8.8cm,height=7.6cm]{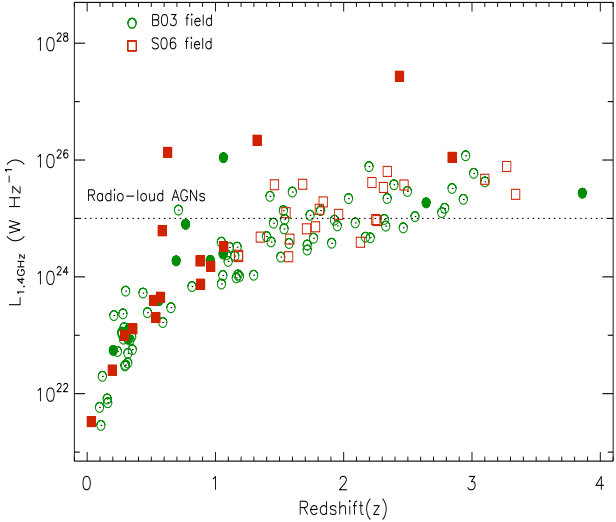}
\caption{Redshift versus 1.4 GHz luminosity plot for our USS sample sources. Green circles and red squares represent
sources in the B03 and the S06 fields, respectively. Filled and open symbols represent sources with the spectroscopic and the photometric redshifts, respectively. 
The dotted line shows the radio$-$loud limit (adopted from \cite{Jiang07,Sajina08}).}
\label{fig:LradioVsZ}
\end{figure}

Radio luminosities of USS sources can be used to infer their possible nature {\ie}radio galaxy, radio$-$quiet AGN, star$-$forming galaxy. 
We study radio luminosity distributions of our USS sample sources.
We use rest-frame radio luminosities that are estimated using k-correction based on spectral index ($\alpha$) measured 
between 325 MHz and 1.4 GHz, and assuming the radio emission is synchrotron emission characterized by a power law 
($S_{\nu}$~$\propto$~${\nu}^{\alpha}$). 
The radio luminosity of a source at redshift $z$ and luminosity-distance $d_{\rm L}$ is therefore given 
by $L_{\nu} = 4\pi d_{\rm L}^{2}$~$S_{\nu} (1+z)^{-({\alpha}+1)}$. 
Figure~\ref{fig:Lradio} shows the 1.4 GHz radio luminosity distributions of our USS sample sources. 
We note that radio luminosities are available only for USS sources with redshift estimates 
{\ie}86/116 sources in the B03 field and 39/44 sources in the S06 field. 
Table~\ref{tab:USSParam} lists the ranges and medians of radio luminosity distributions at 1.4 GHz and 325 MHz of our USS sample sources 
in the two subfields.
Figure~\ref{fig:LradioVsZ} shows the 1.4 GHz radio luminosity versus redshift plot. 
It is clear that most of the low-$z$ ($z < 0.5$) USS sources have 1.4 GHz radio luminosities
(L$_{\rm 1.4~GHz}$) $\sim$ 10$^{21}$ $-$ 10$^{23}$ W Hz$^{-1}$, similar to radio$-$quiet AGNs and star$-$forming galaxies, which is consistent with 
the diagnostics based on the mid-IR colors and the flux ratios of radio to mid-IR. 
We note that a substantially large fraction ({\ie}55/86 $\sim$ 64$\%$ sources in the B03 field, and 31/39 $\sim$ 79.5$\%$ sources in the S06 field) of 
our USS sources do have 1.4 GHz radio luminosity higher than 10$^{24}$ W Hz$^{-1}$. 
Radio sources with L$_{\rm 1.4~GHz}$ $>$ 10$^{24}$ W Hz$^{-1}$ are unlikely to be powered by star formation or starbursts galaxies 
alone ({}\eg\cite{Afonso05}), and likely to constitute radio sources 
such as Compact Steep Spectrum (CSS) radio sources, Gigahertz Peaked Spectrum (GPS) radio sources, and FR$-$I/FR$-$II radio galaxies. 
SMGs with obscured AGN at $z$ $\sim$ 2 $-$ 3, can also have radio luminosities $\sim$ 10$^{24}$ W~Hz$^{-1}$ \citep{Seymour09}.  
Powerful USS radio sources (L$_{\rm 1.4~GHz}$ $>$ 10$^{24}$ W Hz$^{-1}$) with unresolved radio morphologies can be radio sources 
with compact sizes and steep spectra {\ie}CSS and GPS, which are widely thought to represent the start of 
the evolutionary path to large-scale radio sources \citep{Tinti06,Fanti09}.
Majority of our USS sample remain unresolved in our 325 MHz and 1.4 GHz observations (beamsize $\sim$ 6.0 arcsec), and therefore 
high-resolution radio observations are required to determine the morphology, physical extent, and brightness temperature 
of the radio emitting regions and thus allowing us to probe the AGN nature in obscured environments.
In our USS sample, we have a substantial fraction of sources (22/86 $\sim$ 26.6$\%$ sources in the B03 field, and 17/39 $\sim$ 43.6$\%$ sources 
in the S06 field) that do have L$_{\rm 1.4~GHz}$ $\geq$ 10$^{25}$ W Hz$^{-1}$, and can be considered as secure candidate radio$-$loud AGNs ({\eg}\cite{Jiang07,Sajina08}).
Indeed, some of our USS sources 
({\eg}GMRT022735-041121, GMRT022743-042130, GMRT022421-042547, GMRT022733-043317, GMRT022728-040344, GMRT021659-044918, 
GMRT021926-051535, GMRT021827-045440) with L$_{\rm 1.4~GHz}$ $>$ 10$^{25}$ W Hz$^{-1}$, clearly show double-lobed radio morphologies 
at 1.4 GHz, and can be classified as FR$-$I/FR$-$II radio galaxies.

\section{The $K-z$ relation for USS sources} 

\begin{figure}
\centering
\includegraphics[angle=0,width=8.8cm,height=7.6cm]{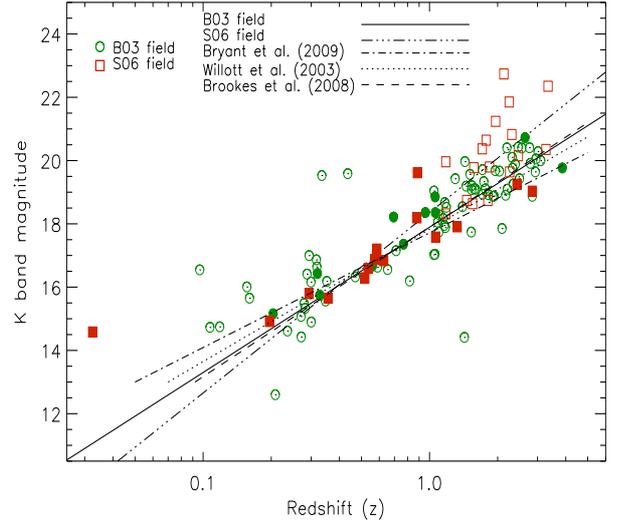}
\caption{$K-z$ plot for our USS sources. Green circles and red squares represent sources in the B03 and the S06 fields, respectively. 
Filled and open symbols represent sources with the spectroscopic and the photometric redshifts, respectively. 
Solid and dashed-triple-dotted lines represent the best fits for sources at redshifts ($z$) $\geq$ 0.5 in the B03 and the S06 fields, respectively.  
The dashed-dotted, dotted and dashed lines represent best fit lines of the $K-z$ relations 
for the powerful radio galaxy samples from \cite{Bryant09,Willott03,Brookes08}, respectively. All the magnitudes are in Vega system.
}
\label{fig:k-zplot}
\end{figure}

It is well known that radio galaxies follow a tight correlation between K-band magnitude and redshift {\ie}$K-z$ relation 
\citep{Jarvis01b,DeBreuck02,Willott03,Brookes08,Bryant09}. 
K-band (centred at 2.2 $\mu$m) observations help to study the stellar population in galaxies over a large redshift range ($0~{\leq}~z~{\leq}~4$) 
as it samples their near-IR to optical rest-frame emission.
We investigate the $K-z$ relation for our USS sample sources. 
The K-band magnitudes in the B03 field and the S06 field are obtained from the VIDEO and the UDS data, respectively. 
The VIDEO magnitudes are in $K_{s}$ band, however, the difference between $K_{s}$ and K band magnitudes is small and 
to the order of typical errors in magnitudes. 
We used Vega magnitudes {\ie}VIDEO K-band AB magnitudes were converted to Vega system using conversion factor (K$_{\rm AB}$ = K$_{\rm Vega}$ + 1.9) 
given in \cite{Hewett06}. 
Some of earlier studies ({\eg}\cite{Eales97,Willott03,DeBreuck04}) used 8.0 arcsec aperture ({\ie}corresponding to 65 kpc at $z$ = 1) K-band magnitude 
to account for the variation of K-band emission with aperture size. 
However, in a sample consists of radio sources with a wide range of flux densities and redshifts, a 4.0 arcsec diameter aperture 
adequately samples nearly the entire K-band emission and reduces the photometric uncertainty (see \cite{Bryant09,Simpson12}). 
Therefore, we use 4.0 arcsec diameter aperture K-band magnitude at all redshifts. 
Also, to find and remove quasars, we performed cross-matching of our USS sample with the SDSS$\footnote{http://www.sdss3.org/}$ DR10 quasar catalog using search radius of 
3.0 arcsec. But, we do not find counterpart of any USS source in the SDSS quasar catalog. 
Therefore, we include all our USS sources in the $K-z$ plot. \\
Figure~\ref{fig:k-zplot} shows the $K-z$ plot for our USS sources with K-band magnitude ranging from 12.0 Mag to 23.0 Mag, and redshift 
spanning over 0.03 to 3.8.  
It is evident that the $K-z$ relation continues to hold for our faint USS sources, although with larger scatter compared to 
the powerful radio galaxies. 
We find that the best linear fits for the USS sources in the B03 and the S06 fields can be represented as $K = 17.89 + 1.99log(z)$, and 
$K = 18.36 + 2.48 log(z)$, respectively, with correlation coefficients 0.73 and 0.61, respectively. 
The best fits and correlation coefficients are obtained by using only sources with redshift ($z$) $\geq$ 0.5 as 
the low-$z$ USS sources are likely to be contaminated by non-AGN star$-$forming galaxies which exhibit larger scatter.  
The comparison of the $K-z$ relation for our USS sample sources with that for powerful radio galaxies {\ie}samples 
from \cite{Willott03}, \cite{Brookes08}, and \cite{Bryant09}, shows that the $K-z$ relation for our faint USS sources is consistent with the 
one seen for bright powerful radio galaxies, however, with a larger scatter.    
The deeper K-band UDS data in the S06 field results in the detection of faint sources at higher redshifts ($z~{\geq}~1.0$). 
These sources tend to deviate from the $K-z$ relation observed for powerful radio galaxies. We note that only photometric redshift estimates are available for these sources. Generally, sources with photometric redshifts tend to show larger scatter than the ones with spectroscopic redshifts and 
therefore, inaccurate photometric redshift estimates may be partly responsible for the larger scatter. 
Also, contamination by AGNs of low radio luminosity can attribute to larger scatter ({\eg}\cite{DeBreuck02,Simpson12}).
Faint radio sources are known to exhibit systematically fainter K-band magnitudes than that for bright radio sources 
at a given redshift ({\eg}\cite{Eales97,Willott03}), which is attributed to different stellar luminosities of their host galaxies. 
Furthermore, in our USS sample, we have a significant fraction ($\sim$ 26$\%$ in the B03 field and $\sim$ 8$\%$ in the S06 field) of sources that 
remained unidentified in the K-band, and these can be considered as potential high-$z$ candidates.

\section{High-$z$ radio sources in faint USS sample}
In our study we use the USS technique to select high-$z$ sources. 
The efficiency of USS technique in selecting high-$z$ sources is based on the existence of the correlation between redshift and radio spectral index 
{\ie}$z$~$-$~$\alpha$ correlation \citep{Klamer06,Ker12}. 
It has been shown that USS samples display higher median redshifts than that for full radio samples ({\eg}\cite{Bryant09}). 
The USS samples selected at different flux limits have achieved varying degree of success in selecting high-$z$ sources. 
There are suggestions that USS method is more efficient in selecting high-$z$ sources at the flux density limit of 
approximately 10 mJy at 1.4 GHz, while the fraction of high-$z$ sources decreases at lower and higher flux densities ({\eg}\cite{Dunlop90,Best03}). 
In the literature, many of the studies on the USS samples have also used additional selection criteria such as small angular size 
and faint infrared magnitude to select high-$z$ sources \citep{DeBreuck04,DeBreuck06,Cruz06}. 
According to the $K-z$ relation, H$z$RGs are expected to be faint in K-band and indeed, 
several H$z$RGs have been discovered by pre-selecting them in K-band \citep{Jarvis01b,Brookes06,Jarvis09}.
The second highest redshift known radio galaxy identified by \cite{Jarvis09} was selected for follow-up based purely on its faint K-band magnitude, 
and it is not a USS source ($\alpha$ = -0.75).   
As discussed in Section 5.2, the radio sources without optical or infrared detections {\ie}infrared-faint radio sources (IFRS) 
also potentially sample high-$z$ sources \citep{Norris11,Middelberg11}. 
Thus, the K-band/IR magnitude based methods are alternative efficient techniques for selecting high-$z$ sources, and can be feasible over large sky area 
with the availability of deep IR and radio surveys. 
A detailed discussion on the comparison of the efficiencies of USS method and K-band/IR based methods to select high-$z$ sources 
is beyond the scope of this paper.
\\
In order to asses how efficient our faint USS sample is in selecting high-$z$ sources, 
we compare the median redshifts and the fraction of faint K-band USS sources in our USS sample to that for well known bright USS samples.
In Table~\ref{tab:USSComparison}, we present a comparison of various parameters 
{\ie}flux limits, USS source densities, median redshifts and the fraction of faint K-band sources in our faint USS sample to that for bright USS samples. 
We find that the USS source density in our sample is nearly 1000 times higher than that for the bright samples 
{\eg}6C$^{\star}$ \citep{Jarvis01b}, SUMSS-NVSS \citep{DeBreuck04}, WENSS-NVSS \citep{DeBreuck2000,DeBreuck02}. 
This can understood as we are probing at sub-mJy regime which is two order of magnitude deeper than the bright USS sample 
flux limits {\ie}10 - 15 mJy in shallow and wide area surveys. 
The comparison of median redshifts shows that our faint USS sample has median redshift similar to the one for SUMSS-NVSS sample, although, 6C$^{\star}$, 
WENSS-NVSS samples do have higher median redshifts. 
It is to be noted that the bright samples have additional biases due to K-band selection and incomplete spectroscopic redshifts, 
and hence a direct comparison may not be viable, but it is interesting to note that the median redshifts are broadly consistent. 
In Table~\ref{tab:USSComparison}, we also present the comparison of the fraction of USS sources with K-band magnitude fainter than 19.5 Vega magnitude (K $>$ 19.5). 
The fraction of faint K-band sources in the sample can be used as an indicator of the fraction of high-$z$ sources owing to the $K-z$ relation. 
We note that the fraction of USS sources with K $>$ 19.5 in our sample is $\sim$ 30$\%$, similar to one found in the bright USS samples {\eg}WENSS-NVSS, TEXAS-NVSS 
and SUMSS-NVSS. 
Moreover, K-band photometry of the WENSS-NVSS sample is not complete as most of the sources observed in K-band were pre-selected to be those 
which were not detected in optical imaging. 
This kind of pre-selection is likely to remove a significant fraction of intermediate redshift sources with K $<$ 19.5. 
\\
In conclusion, we state that the comparison of the median redshifts and the fractions of USS sources with faint K-band magnitudes of our sample with 
that of the bright USS samples, suggests that even at faint flux density, the USS selection is an efficient method to select high-$z$ sources. 
The high-$z$ USS sources in our sample do have faint optical/IR counterparts ({\cf}Sections 3 and 5) and this may be 
the combined effect of the $z-{\alpha}$ correlation and the $K-z$ correlation. 
Our study on the faint USS sources limited to small sky area ({\ie}1.8 deg$^{-2}$) can be used as the basis to search for 
high-$z$ sources via USS technique in the next generation wide and deep radio continuum surveys down to $\mu$Jy level 
{\eg}from SKA pathfinders \citep{Norris11b,Norris13} and LOFAR \citep{VanHaarlem13}. 
The deep optical/IR follow up surveys ({\eg}from LSST \citep{Ivezic08}, JWST \citep{Gardner06}, WFIRST \citep{Green12}) 
will help us in obtaining photometric redshifts and in removing low redshift contaminants.

\begin{table*} 
\caption{Fraction of high-z sources in faint and bright USS samples}
\begin{minipage}{140mm}
\begin{tabular} {ccccccccc}
\hline
Sample     & Flux limit & Spectral index  &  Area       &  Sources  &  USS density  &  Median   & Fraction of USS    & Ref. \\ 
           & (mJy)      & limit           &  sr         &           &  (sr$^{-1}$)  &  Redshift &  with K $>$ 19.5     &      \\  \hline
WENSS-NVSS & S$_{\rm 1.4~GHz}$ $>$ 10  & ${\alpha}_{\rm 325~MHz}^{\rm 1.4~GHz}$ $<$ -1.3  &  2.27 & 343  & 151  &  1.87  & 12/44 (27$\%$) & 1  \\ 
TEXAS-NVSS & S$_{\rm 1.4~GHz}$ $>$ 10  & ${\alpha}_{\rm 365~MHz}^{\rm 1.4~GHz}$ $<$ -1.3  &  5.58 & 268  & 48   &  2.10  & 8/24 (33$\%$) & 1  \\ 
MRC-PMN    & S$_{\rm 408~MHz}$ $>$ 700 & ${\alpha}_{\rm 408~MHz}^{\rm 4.8~GHz}$ $<$ -1.2  &  2.23 &  58  & 26   &  0.88  & 0/29  & 1   \\ 
6C*        & S$_{\rm 151~MHz}$ $>$ 960 & ${\alpha}_{\rm 151~MHz}^{\rm 4.8~GHz}$ $<$ -0.981&  0.133&  29  & 218  &  1.90  & 2/24 (8$\%$) & 2   \\ 
SUMSS-NVSS & S$_{\rm 1.4~GHz}$ $>$ 15  & ${\alpha}_{\rm 843~MHz}^{\rm 1.4~GHz}$ $<$ -1.3  &  0.11 &  53  & 482  &  1.20  & 13/53 (25$\%$) & 3   \\ 
$^{\dagger}$VLA-GMRT   & S$_{\rm 610~MHz}$ $>$ 0.1 & ${\alpha}_{\rm 610~MHz}^{\rm 1.4~GHz}$ $<$ -1.3& 1.71 $\times$ 10$^{-4}$   &  58  & 3.40 $\times$ 10$^{5}$  &  0.60  & .... &  4    \\
Our        & S$_{\rm 325~MHz}$ $>$ 0.5 & ${\alpha}_{\rm 325~MHz}^{\rm 1.4~GHz}$ $\leq$ -1.0& 5.48 $\times$ 10$^{-4}$ & 160 & 2.92 $\times$ 10$^{5}$ & 1.31 & 35/117 (30$\%$) & 5 \\ \hline 
\end{tabular} 
\label{tab:USSComparison} 
\vspace {0.05cm} \\
Notes - References: (1) \cite{DeBreuck2000,DeBreuck02}; (2) \cite{Blundell98,Jarvis01a}; (3) \cite{DeBreuck04,DeBreuck06}; 
(4) \cite{Afonso11}; (5) this paper. \\
The comparison of flux limits, USS source densities for bright samples is given in \cite{DeBreuck04}. \\
$\dagger$ : K-band magnitudes are unavailable for the USS sample sources (${\alpha}_{\rm 610~MHz}^{\rm 1.4~GHz}$ $<$ -1.3) presented by \cite{Afonso11}.
The median redshift of sample is based on only sources with available redshift estimates. 
We opted average mean value of redshift, if median redshift is unavailable for bright USS sample. 
Low median redshift for faint USS sample of \cite{Afonso11} is likely to be the result of the unavailability of redshift 
of 47$\%$ sample sources that are faint at 3.6 $\mu$m and are candidate high-$z$ sources.    
\end{minipage}
\end{table*}

\section{Conclusions}
Using the most sensitive 325 MHz GMRT observations (5$\sigma$ $\sim$ 800 $\mu$Jy) and 1.4 GHz VLA 
observations (5$\sigma$ $\sim$ 80 - 100 $\mu$Jy) available for two subfields in the XMM-LSS field,
we derive a large sample of 160 faint USS sources (${\alpha}_{\rm 325~MHz}^{\rm 1.4~GHz}$ $\leq$ -1). 
Our study is one of the few attempts made in the literature to characterize the population of faint USS sources down to sub-mJy level, 
and to search for H$z$RG candidates. 
The availability of deep optical, near-IR data in the two subfields allow us to identify counterparts of the majority of our USS sample sources, 
and to unveil their nature.
The conclusions of our study are:
\begin{enumerate}
\item Using the CFHTLS-D1 optical data (r$^{\prime}_{\rm AB}$ $\sim$ 26.1) in the B03 field, and Subaru/SuprimeCam data 
(R$_{\rm AB}$ $\sim$ 27.7) in the S06 field, we find optical counterparts of 86/116 $\sim$ 74$\%$ 
and 37/39 $\sim$ 95$\%$ USS sources in the two subfields, respectively. 
In near-IR, the VIDEO data (K$_{\rm AB}$ $\sim$ 23.5), and the UDS data (K$_{\rm AB}$ $\sim$ 24.6), 
yield similar high identification rates {\ie}86/116 $\sim$ 74$\%$ and 35/38 $\sim$ 92$\%$ in the B03 and the S06 fields, respectively. 
The Spitzer surveys at 3.6 $\mu$m and 4.5 $\mu$m {\ie}the SERVS data ([3.6]$_{\rm AB}$ $\sim$ 23.1) in the B03 field, 
and SpUDS data ([3.6]$_{\rm AB}$ $\sim$ 24.0) in the S06 field yield counterparts for 72/95 $\sim$ 76$\%$ 
and 32/36 $\sim$ 89$\%$ USS sources in the two subfields, respectively ({\cf}Table~\ref{table:OpticalMags}). 
We find that, in compared to full radio population, the optical and IR magnitude distributions of USS sources are systematically flatter and 
fainter. This can be interpreted as the possible dusty and/or high-$z$ nature of USS sources.

\item 
Redshift estimates are available for 86/116 $\sim$ 74$\%$ and 39/44 $\sim$ 89$\%$ of the USS sources in the B03 and the S06 field, respectively. 
The distributions of available redshifts for our USS sample sources span over $z$ $\sim$ 0.03 to 3.86 with 
the median values $z_{\rm median}$ $\sim$ 1.18, and $z_{\rm median}$ $\sim$ 1.57 in the B03 and the S06 fields, respectively. 
The lower median redshift in the B03 field can be attributed to the fact that the redshift estimates are not available for a large fraction 
(30/116 $\sim$ 26$\%$) of USS sources, and the radio to mid-IR flux ratio diagnostic suggests these to be potentially high redshift candidates. 
Also, the USS sources show higher median redshifts than that for the non-USS radio sources. 
The comparison of the redshift distributions of our USS sources with the one for the radio population derived by using the 
SKADS Simulated Skies (S$^{3}$) simulations, shows that our faint USS sample efficiently selects high-$z$ sources. 
However, due to faint flux density limit our USS sample may be dominated by the less powerful radio-loud sources. 

\item The mid-IR color-color diagnostics are consistent with majority of our USS sample sources at relatively higher redshifts ($z > 0.5$) being 
mainly AGN. However, at low redshift ($z< 0.5$) USS sample may contain sources in which mid-IR colors are dominated by the emission due to star formation. 

\item A substantially large fraction of our USS sources (nearly 33/88 $\sim$ 38$\%$ in the B03 field and 21/36 $\sim$ 58$\%$ in 
the S06 field) have radio to mid-IR flux ratios (S$_{\rm 1.4~GHz}$/S$_{\rm 3.6~{\mu}m}$) $\sim$ 4 to 100, distributed over $z$ ${\sim}$ 0.5 to 3.8. 
The locations {\bf of} these USS sources in the radio to mid-IR flux ratio (S$_{\rm 1.4~GHz}$/S$_{\rm 3.6~{\mu}m}$) versus redshift 
diagnostic plot is similar to the one observed for Sub-Millimeter Galaxies (SMGs) in the representative sample of Norris et al. (2011). 
The radio luminosities (L$_{\rm 1.4~GHz}$ $\geq$ 10$^{24}$ W Hz$^{-1}$) and compact radio sizes 
suggest these USS sources to be potentially weakly radio$-$loud AGN hosted in obscured environments.

\item  There are 23/88 $\sim$ 26$\%$ USS source in the B03 field, and 4/36 $\sim$ 11$\%$ USS source in the S06 field that do not have 
3.6 $\mu$m detection and exhibit high radio to mid-IR flux ratio limits {\ie}S$_{\rm 1.4~GHz}$/S$_{\rm 3.6~{\mu}m}$ $>$ 50.
The flux ratios of these USS sources are similar to the ones observed for radio$-$loud AGNs and powerful H$z$RGs, and therefore, these 
USS sources can be considered as H$z$RGs candidates.

\item Radio luminosity distributions of our USS sources span over wide range 
{\eg}L$_{\rm 1.4~GHz}$ $\sim$ 10$^{21}$ W Hz$^{-1}$ $-$ 10$^{27}$ W Hz$^{-1}$. 
A significant fraction of our USS sources ({\ie}22/86 $\sim$ 26.6$\%$ sources in the B03 field, and 17/39 $\sim$ 43.6$\%$ sources in the S06 field), 
do have L$_{\rm 1.4~GHz}$ $\geq$ 10$^{25}$ W Hz$^{-1}$, and can be considered as secure candidate radio$-$loud AGNs.
USS sources of high radio luminosities (L$_{\rm 1.4~GHz}$ $>$ 10$^{24}$ W Hz$^{-1}$) with unresolved radio morphologies 
can be sources with compact sizes and steep spectra {\eg}CSS and GPS, which are thought to represent the start of 
the evolutionary path to large-scale radio sources \citep{Tinti06,Fanti09}. 
However, high-resolution radio observations are required to determine the morphology, physical extent, and brightness temperature 
of the radio emitting regions and probe their nature. 

\item 
Our USS sources follow the $K-z$ relation, although with larger scatter compared to powerful radio galaxies.
The comparison of $K-z$ relation of our USS sources with the one for H$z$RGs suggests that apart from H$z$RG candidates our USS sample also 
contain radio sources of various classes such as weakly radio$-$loud sources at higher redshifts and radio$-$quiet AGNs at low redshift.  
\item 
Our study demonstrates that the criterion of ultra steep spectral index remains an efficient method to select high redshift sources 
even at sub-mJy flux densities. We find that, in addition to powerful H$z$RG candidates, faint USS population also contain weak radio$-$loud 
AGNs likely to be hosted in obscured environments. In our forthcoming paper we shall investigate the nature of obscured environments of these sources 
using far-IR/sub-mm observations from Herschel. 
 
\end{enumerate}

\begin{acknowledgements}
We gratefully acknowledge generous support from the Indo-French Center for the Promotion of Advanced Research (Centre Franco-Indien pour la 
Promotion de la Recherche Avance) under program no. 4404-3.
We thank the staff of GMRT who have made these observations possible. 
GMRT is run by the National Centre for Radio Astrophysics of the Tata Institute of Fundamental Research. 
We thank Marco Bondi for providing 1.4 GHz and 610 MHz radio images of VVDS field.
We thank Chris Simpson for providing 1.4 GHz VLA radio image of SXDF field. 
We also thank the anonymous referee for useful comments which helped to improve the manuscript. 
This work is based on observations made with the Spitzer Space Telescope, which is operated by the Jet
Propulsion Laboratory (JPL), California Institute of Technology (Caltech), under a contract with NASA.
This work used the CFHTLS data products, which are based on observations obtained with MegaPrime/MegaCam, a joint project of
CFHT and CEA/DAPNIA, at the CFHT which is operated by the National Research Council (NRC) of Canada, the Institut National
des Science de l'Univers of the Centre National de la Recherche Scientifique (CNRS) of France, and the University of Hawaii. 
This work is based in part on data products produced at TERAPIX and the Canadian Astronomy Data Centre as part of the CFHTLS, a
collaborative project of NRC and CNRS. 
This research uses data from the VIMOS VLT Deep Survey, obtained from the VVDS database operated by 
Cesam, Laboratoire d'Astrophysique de Marseille, France. 
\end{acknowledgements}

\newpage

\begin{appendix}
\section{Our USS Sample}
In table A.1 we list all our USS sample sources derived from 325 MHz and 1.4~GHz observations.

\begin{table*}
\centering
\begin{minipage}{140mm}
\caption{USS sample}
\small
\begin{tabular}{@{}ccccccccc@{}}
\hline
Source    & RA$_{\rm 1.4~GHz}$ & DEC$_{\rm 1.4~GHz}$ &   S$_{\rm 325~MHz}$ & S$_{\rm 1.4~GHz}$ &  ${\alpha}_{\rm 325 MHz}^{\rm 1.4 GHz}$ & $z_{\rm phot}$ & $z_{\rm spec}$ & logL$_{\rm 1.4~GHz}$  \\
 name     & (hms) &  (dms) &   (mJy)             &  (mJy)            &                                         &                &                &   (W Hz$^{-1}$)       \\ \hline
{\bf S06 field}   &             &             &                 &                 &                &       &      &       \\
  GMRT021611-050101 & 02 16 11.70 & -05 00 53.54 & 3.441$\pm$0.607 & 0.150$\pm$0.017 & -2.15$\pm$0.14 & 3.27 & ....  & 25.89\\
  GMRT021618-050522 & 02 16 18.99 & -05 05 18.87 & 1.050$\pm$0.195 & 0.181$\pm$0.018 & -1.20$\pm$0.14 & 2.26 & ....  & 24.96\\
  GMRT021620-045923 & 02 16 20.33 & -04 59 21.75 & 6.873$\pm$0.120 & 1.522$\pm$0.018 & -1.03$\pm$0.01 & 2.32 & 2.845 & 26.05\\
  GMRT021635-050651 & 02 16 34.57 & -05 06 48.17 & 1.937$\pm$0.114 & 0.341$\pm$0.015 & -1.19$\pm$0.05 & 1.96 & ....  & 25.07 \\
  $^{\star}$GMRT021646-051004 & 02 16 46.93 & -05 10 01.94 & 1.928$\pm$0.118 & 0.423$\pm$0.022 & -1.04$\pm$0.05 & .... & .... & .... \\
  $^{\star}$GMRT021648-045838 & 02 16 48.61 & -04 58 43.26 & 0.551$\pm$0.104 & 0.086$\pm$0.021 & -1.27$\pm$0.21 & .... & .... & .... \\
  GMRT021649-051859 & 02 16 49.46 & -05 18 57.70 & 1.080$\pm$0.124 & 0.197$\pm$0.014 & -1.16$\pm$0.09 & 3.34 & ....  & 25.41\\
  GMRT021656-053001 & 02 16 56.55 & -05 30 00.21 & 2.855$\pm$0.168 & 0.534$\pm$0.017 & -1.15$\pm$0.05 & 1.81 & .... & 25.16\\
$^{\dagger}$GMRT021659-044918 & 02 16 59.02 & -04 49 20.53 & 155.26$\pm$1.95 & 9.600$\pm$0.135 & -1.91$\pm$0.01 & 1.31 & 1.325& 26.34\\
  GMRT021702-045721 & 02 17 02.49 & -04 57 19.69 & 3.472$\pm$0.157 & 0.784$\pm$0.014 & -1.02$\pm$0.03 & 1.84 & ....& 25.28\\
  GMRT021706-044705 & 02 17 06.29 & -04 47 04.67 & 0.850$\pm$0.104 & 0.193$\pm$0.014 & -1.02$\pm$0.10 & 0.84 & 0.884& 23.88\\
  GMRT021713-050638 & 02 17 13.55 & -05 06 41.07 & 1.485$\pm$0.292 & 0.286$\pm$0.013 & -1.13$\pm$0.14 & 1.78 & ....& 24.86\\
 $^{\star}$GMRT021648-045838 & 02 16 48.61 & -04 58 43.26 & 0.551$\pm$0.104 & 0.086$\pm$0.021 & -1.27$\pm$0.21 & .... & .... & ....\\
  GMRT021716-045140 & 02 17 16.67 & -04 51 40.21 & 1.962$\pm$0.104 & 0.155$\pm$0.031 & -1.74$\pm$0.14 & .... & .... &  .... \\
  GMRT021718-053206 & 02 17 18.18 & -05 32 06.37 & 3.287$\pm$0.199 & 0.708$\pm$0.016 & -1.05$\pm$0.04 & 2.47 & .... & 25.57\\
  GMRT021723-043515 & 02 17 23.82 & -04 35 13.72 & 2.372$\pm$0.177 & 0.360$\pm$0.018 & -1.29$\pm$0.06 & 3.10 & .... & 25.67\\
  GMRT021725-051620 & 02 17 25.11 & -05 16 17.27 & 0.601$\pm$0.131 & 0.132$\pm$0.012 & -1.04$\pm$0.16 & 1.57 & .... & 24.34\\
  GMRT021725-044130 & 02 17 25.89 & -04 41 30.78 & 0.506$\pm$0.107 & 0.101$\pm$0.013 & -1.10$\pm$0.17 & 2.13 & .... & 24.59\\
  GMRT021726-051428 & 02 17 25.98 & -05 14 26.93 & 1.074$\pm$0.114 & 0.218$\pm$0.013 & -1.09$\pm$0.08 & 2.25 & .... & 24.98\\
  GMRT021734-051957 & 02 17 34.39 & -05 19 56.45 & 1.476$\pm$0.149 & 0.311$\pm$0.043 & -1.07$\pm$0.12 & 1.71 & .... & 24.82\\
  GMRT021740-045148 & 02 17 40.69 & -04 51 44.21 & 8.843$\pm$0.691 & 0.195$\pm$0.013 & -2.61$\pm$0.07 & 0.42 & 0.518& 23.60\\
 $^{\star}$GMRT021742-045842 & 02 17 42.67 & -04 58 38.46 & 0.585$\pm$0.114 & 0.076$\pm$0.021 & -1.40$\pm$0.23 & .... & .... & .... \\
  GMRT021743-051748 & 02 17 43.84 & -05 17 51.45 & 6.324$\pm$0.390  & 1.410$\pm$0.049 & -1.03$\pm$0.05 & 0.03 & 0.033& 21.52\\
  GMRT021743-052810 & 02 17 44.07 & -05 28 09.20 & 1.365$\pm$0.169 & 0.273$\pm$0.015 & -1.10$\pm$0.09 & 1.18 & .... & 24.37\\
  GMRT021745-050057 & 02 17 45.84 & -05 00 56.41 & 5.420$\pm$0.109 & 0.590$\pm$0.013 & -1.52$\pm$0.02 & 2.22 & .... & 25.62\\
  GMRT021754-051250 & 02 17 54.10 & -05 12 49.94 & 21.13$\pm$0.14 & 4.200$\pm$0.061 & -1.11$\pm$0.01 & 0.51 & 0.586& 24.79\\
  GMRT021800-051147 & 02 18 00.52 & -05 11 44.76 & 1.502$\pm$0.124 & 0.295$\pm$0.013 & -1.11$\pm$0.06 & 0.29 & 0.356& 23.11\\
  GMRT021800-053602 & 02 18 00.82 & -05 36 01.75 & 1.278$\pm$0.179 & 0.226$\pm$0.017 & -1.19$\pm$0.11 & 1.58 & .... & 24.65\\
  GMRT021803-044745 & 02 18 03.08 & -04 47 41.83 & 1.623$\pm$0.219 & 0.325$\pm$0.036 & -1.10$\pm$0.12 & 0.48 & 0.572& 23.65\\
  GMRT021803-043912 & 02 18 03.29 & -04 39 11.71 & 2.425$\pm$0.226 & 0.500$\pm$0.014 & -1.08$\pm$0.07 & 0.99 & 1.064& 24.51\\
  GMRT021811-053236 & 02 18 11.16 & -05 32 34.31 & 1.408$\pm$0.191 & 0.236$\pm$0.015 & -1.22$\pm$0.10 & 1.18 & .... & 24.35\\
 $^{\star}$GMRT021814-051456 & 02 18 14.35 & -05 14 53.74 & 3.579$\pm$0.111 & 0.683$\pm$0.250 & -1.13$\pm$0.25 & .... & .... & .... \\
 $^{\dagger}$GMRT021827-045440 & 02 18 27.32 & -04 54 37.29 &367.94$\pm$0.71  & 80.25$\pm$0.07 & -1.04$\pm$0.01 & 0.58 & 0.627& 26.13\\
  GMRT021830-050100 & 02 18 30.65 & -05 00 55.58 & 2.601$\pm$0.281 & 0.419$\pm$0.038 & -1.25$\pm$0.10 & 0.87 & 0.88 & 24.27\\
  GMRT021830-050421 & 02 18 30.28 & -05 04 20.34 & 0.901$\pm$0.147 & 0.168$\pm$0.012 & -1.15$\pm$0.12 & 0.44 & 0.536& 23.30\\
  GMRT021831-053632 & 02 18 31.38 & -05 36 31.22 & 1.939$\pm$0.197 & 0.406$\pm$0.020 & -1.07$\pm$0.08 & 1.35 & .... & 24.68\\
  GMRT021838-053445 & 02 18 38.29 & -05 34 44.98 & 9.712$\pm$0.236 & 1.580$\pm$0.019 & -1.24$\pm$0.02 & 1.68 & .... & 25.58\\
  GMRT021839-044150 & 02 18 39.53 & -04 41 50.10 &250.15$\pm$0.44  & 50.82$\pm$0.07  & -1.09$\pm$0.01 & 2.18 & 2.435& 27.43\\
  GMRT021847-052811 & 02 18 47.22 & -05 28 11.81 & 3.484$\pm$0.232 & 0.797$\pm$0.052 & -1.01$\pm$0.06 & 2.31 & .... & 25.53\\
  GMRT021849-052159 & 02 18 49.79 & -05 21 57.89 & 1.653$\pm$0.175 & 0.368$\pm$0.015 & -1.03$\pm$0.08 & 0.23 & 0.294& 23.00\\
  GMRT021908-051637 & 02 19 08.39 & -05 16 36.04 & 3.460$\pm$0.165 & 0.801$\pm$0.016 & -1.00$\pm$0.04 & 1.55 & .... & 25.10\\
  GMRT021912-050503 & 02 19 12.43 & -05 05 01.55 & 1.512$\pm$0.332 & 0.223$\pm$0.037 & -1.31$\pm$0.19 & 0.20 & 0.197& 22.41\\
$^{\dagger}$GMRT021926-051535 & 02 19 26.48 & -05 15 35.00 & 13.29$\pm$0.38 & 2.390$\pm$0.092 & -1.17$\pm$0.03 & 1.46 & .... & 25.58\\
  GMRT021942-050727 & 02 19 41.90 & -05 07 27.80 & 1.488$\pm$0.168 & 0.300$\pm$0.017 & -1.10$\pm$0.09 & 0.91 & 0.963& 24.18\\
  GMRT021945-045623 & 02 19 45.72 & -04 56 19.45 & 6.602$\pm$0.212 & 1.216$\pm$0.051 & -1.16$\pm$0.04 & 2.34 & .... & 25.81\\
{\bf B03 field}   &             &             &                 &                 &                &       &      &       \\
  GMRT022404-043520 & 02 24 04.14 & -04 35 20.40 & 0.695$\pm$0.131 & 0.109$\pm$0.019 & -1.27$\pm$0.17 & 1.72 & .... & 24.46\\
 $^{\star}$GMRT022405-043553 & 02 24 05.09 & -04 35 53.30 & 1.050$\pm$0.132 & 0.052$\pm$0.016 & -2.06$\pm$0.23 & .... & .... & \\
  GMRT022410-042240 & 02 24 10.09 & -04 22 36.0  & 0.666$\pm$0.116 & 0.120$\pm$0.018 & -1.17$\pm$0.16 & 1.175 & .... & 24.03\\
  GMRT022410-044608 & 02 24 10.13 & -04 46 07.5  & 85.60$\pm$0.27  & 18.89$\pm$0.02  & -1.03$\pm$0.01 & .... & .... & .... \\
  GMRT022410-042156 & 02 24 09.98 & -04 21 47.6  & 3.589$\pm$0.199 & 0.418$\pm$0.024 & -1.47$\pm$0.05 & 2.037 & .... & 25.34\\
$^{\star}$GMRT022411-040004 & 02 24 11.37 & -04 00 00.0  & 1.524$\pm$0.174 & 0.215$\pm$0.037 & -1.34$\pm$0.14 & .... & .... & ....\\
  GMRT022412-041043 & 02 24 12.24 & -04 10 41.4  & 1.771$\pm$0.137 & 0.392$\pm$0.016 & -1.03$\pm$0.06 & 1.398 & .... & 24.69\\
  GMRT022412-044044 & 02 24 12.32 & -04 40 43.0  & 1.729$\pm$0.217 & 0.268$\pm$0.020 & -1.28$\pm$0.10 & 0.470 & .... & 23.39\\
  GMRT022412-045314 & 02 24 12.70 & -04 53 13.7  & 0.773$\pm$0.116 & 0.146$\pm$0.018 & -1.14$\pm$0.13 & 0.318 & .... & 22.69\\
  GMRT022413-044643 & 02 24 13.15 & -04 46 42.0  & 4.017$\pm$0.149 & 0.327$\pm$0.020 & -1.72$\pm$0.05 & 1.453 & .... & 24.92\\
\hline
\end{tabular}
\label{tab:USSSample} 
\vspace {0.05cm} 
Notes- Column 1: Source name as given in our 325 MHz GMRT catalog (Sirothia et al. 2014, in preparation); 
Column 2 and Column 3: RA and DEC from 1.4 GHz VLA observations presented in \cite{Bondi03} and \cite{Simpson06}; 
Column 4: 325 MHz flux density in mJy from our GMRT observations; Column 5: 1.4 GHz flux density in mJy from \cite{Bondi03} and \cite{Simpson06};
Column 6: Radio spectral index between 325 MHz to 1.4 GHz; Column 7: Photometric redshift estimates taken from \cite{McAlpine13} 
and \cite{Simpson12} for the sources in the B03 and S06 fields, respectively; Column 8: Spectroscopic redshifts taken from VVDS catalog 
\citep{LeFevre13} and \cite{Simpson12}; Column 9: 1.4 GHz radio luminosity in logarithms. 
Spectroscopic redshifts, whenever available, are given preference over photometric redshifts to estimate the radio luminosities.
`$^{\star}$' indicates sources detected at $<$5$\sigma$ in 1.4~GHz, while `$^{\dagger}$' indicates sources with clear double lobe radio morphology. \\
{\it Continuing on next page}. 
\end{minipage}
\end{table*}

\addtocounter{table}{-1}

\begin{table*}
\centering
\begin{minipage}{140mm}
\caption{USS sample}
\small
\begin{tabular}{@{}ccccccccc@{}}
\hline
Source    & RA$_{\rm 1.4~GHz}$ &  DEC$_{\rm 1.4~GHz}$ &   S$_{\rm 325~MHz}$ & S$_{\rm 1.4~GHz}$ &  ${\alpha}_{\rm 325 MHz}^{\rm 1.4 GHz}$ & $z_{\rm phot}$ & $z_{\rm spec}$ & logL$_{\rm 1.4~GHz}$  \\
 name     & (hms) &  (dms) &   (mJy)             &  (mJy)            &                                         &                &                &   (W Hz$^{-1}$)       \\ \hline
 $^{\star}$GMRT022414-043242 & 02 24 14.58 & -04 32 41.3  & 0.631$\pm$0.102 & 0.136$\pm$0.042 & -1.05$\pm$0.24 & .... & ....&  ....\\
  GMRT022416-042401 & 02 24 16.77 & -04 24 00.1  & 0.607$\pm$0.118 & 0.117$\pm$0.018 & -1.13$\pm$0.17 & 1.047 & .... & 23.88\\
$^{\dagger}$GMRT022421-042547 & 02 24 20.96 & -04 25 44.6  &108.78$\pm$0.35 & 18.97$\pm$0.03 & -1.20$\pm$0.01 & .... &  .... & .... \\
  GMRT022425-044828 & 02 24 25.78 & -04 48 29.4  & 2.177$\pm$0.225 & 0.335$\pm$0.016 & -1.28$\pm$0.08 & 1.169 & .... & 24.51\\
  GMRT022425-042738 & 02 24 25.95 & -04 27 36.8  & 1.170$\pm$0.148 & 0.099$\pm$0.016 & -1.69$\pm$0.14 & 2.316 & .... & 24.98\\
  GMRT022427-045933 & 02 24 27.54 & -04 59 33.0  & 2.557$\pm$0.161 & 0.509$\pm$0.023 & -1.11$\pm$0.05 & 0.288 & .... & 23.13\\
  GMRT022430-040441 & 02 24 30.12 & -04 04 38.4  & 0.531$\pm$0.107 & 0.085$\pm$0.014 & -1.25$\pm$0.18 & ....  & .... & ....\\
  GMRT022430-045331 & 02 24 30.60 & -04 53 26.6  & 1.745$\pm$0.221 & 0.199$\pm$0.015 & -1.49$\pm$0.10 & 1.434 & .... & 24.60\\
  GMRT022432-041341 & 02 24 32.18 & -04 13 42.9  & 1.312$\pm$0.173 & 0.250$\pm$0.016 & -1.14$\pm$0.10 & 1.099 & .... & 24.26\\
  GMRT022433-043709 & 02 24 33.12 & -04 37 07.1  & 2.470$\pm$0.113 & 0.415$\pm$0.017 & -1.22$\pm$0.04 & ....  & .... & .... \\
  GMRT022433-040748 & 02 24 33.47 & -04 07 48.9  & 0.657$\pm$0.120 & 0.142$\pm$0.015 & -1.05$\pm$0.14 & 1.509 & .... & 24.34\\
  GMRT022436-041046 & 02 24 36.28 & -04 10 50.5  & 2.818$\pm$0.289 & 0.288$\pm$0.015 & -1.56$\pm$0.08 & 0.236 & .... & 22.72\\
  GMRT022436-041007 & 02 24 37.00 & -04 10 01.5  & 1.852$\pm$0.165 & 0.334$\pm$0.020 & -1.17$\pm$0.07 & 0.282 & .... & 22.93\\
  GMRT022444-042658 & 02 24 43.90 & -04 26 51.5  & 1.170$\pm$0.188 & 0.253$\pm$0.024 & -1.05$\pm$0.13 & 0.097 & .... & 21.76\\
  GMRT022447-042948 & 02 24 47.35 & -04 29 44.1  & 0.509$\pm$0.108 & 0.116$\pm$0.015 & -1.01$\pm$0.17 & 0.59  & .... & 23.22\\
  GMRT022447-045436 & 02 24 47.51 & -04 54 33.6  & 3.635$\pm$0.131 & 0.583$\pm$0.015 & -1.25$\pm$0.03 & ....  & .... & .... \\
  GMRT022449-045034 & 02 24 50.01 & -04 50 32.2  & 0.537$\pm$0.103 & 0.121$\pm$0.015 & -1.02$\pm$0.16 & 2.207 & .... & 24.67\\
  GMRT022454-040628 & 02 24 54.80 & -04 06 31.9  & 0.579$\pm$0.117 & 0.124$\pm$0.016 & -1.06$\pm$0.16 & 2.169 & .... & 24.68\\
  GMRT022458-042601 & 02 24 58.59 & -04 26 01.9  & 3.408$\pm$0.195 & 0.414$\pm$0.016 & -1.44$\pm$0.05 & 1.540 & .... & 24.98\\
  GMRT022508-040650 & 02 25 08.35 & -04 06 52.7  & 3.922$\pm$0.222 & 0.737$\pm$0.016 & -1.14$\pm$0.04 & 0.435 & .... & 23.72\\
  GMRT022508-043829 & 02 25 08.54 & -04 38 25.0  & 0.708$\pm$0.110 & 0.132$\pm$0.017 & -1.15$\pm$0.14 & 0.352 & .... & 22.75\\
  GMRT022509-040103 & 02 25 09.18 & -04 01 01.6  & 27.15$\pm$0.15  & 6.098$\pm$0.017 & -1.02$\pm$0.01 & 0.712 & .... & 25.14\\
  GMRT022509-044653 & 02 25 09.57 & -04 46 51.0  & 0.484$\pm$0.114 & 0.103$\pm$0.018 & -1.06$\pm$0.20 & 0.317 & .... & 22.53\\
  GMRT022510-040403 & 02 25 10.22 & -04 04 00.6  & 6.668$\pm$0.125 & 1.505$\pm$0.018 & -1.02$\pm$0.02 & 2.950 & .... & 26.07\\
  GMRT022517-041755 & 02 25 18.03 & -04 17 53.2  & 3.117$\pm$0.127 & 0.620$\pm$0.022 & -1.11$\pm$0.04 & 1.046 & .... & 24.59\\
  GMRT022517-042410 & 02 25 18.15 & -04 24 07.8  & 0.660$\pm$0.106 & 0.100$\pm$0.017 & -1.29$\pm$0.16 & ....  & .... & .... \\
  GMRT022519-042754 & 02 25 19.84 & -04 27 51.7  & 0.975$\pm$0.103 & 0.125$\pm$0.016 & -1.41$\pm$0.11 & ....  & .... & .... \\
  GMRT022525-044641 & 02 25 26.50 & -04 46 40.6  & 10.85$\pm$0.59  & 1.946$\pm$0.080 & -1.18$\pm$0.05 & ....  & .... & .... \\
  GMRT022526-042119 & 02 25 26.95 & -04 21 17.3  & 0.777$\pm$0.115 & 0.169$\pm$0.018 & -1.04$\pm$0.12 & 1.058 & .... & 24.02\\
  GMRT022528-041535 & 02 25 28.05 & -04 15 36.3  & 1.764$\pm$0.261 & 0.278$\pm$0.024 & -1.27$\pm$0.12 & 0.709 &0.559& 23.59\\
  GMRT022530-043936 & 02 25 30.33 & -04 39 35.4  & 5.377$\pm$0.136 & 1.146$\pm$0.024 & -1.06$\pm$0.02 & ....  & .... & ....\\
  GMRT022534-042248 & 02 25 34.55 & -04 22 43.7  & 0.570$\pm$0.093 & 0.099$\pm$0.017 & -1.20$\pm$0.16 & 0.160 & .... & 21.84\\
  GMRT022534-044543 & 02 25 34.85 & -04 45 42.0  & 1.297$\pm$0.094 & 0.269$\pm$0.017 & -1.08$\pm$0.07 & ....  & .... & ....\\
  GMRT022536-042146 & 02 25 36.30 & -04 21 42.7  & 3.626$\pm$0.202 & 0.540$\pm$0.021 & -1.30$\pm$0.05 & 0.118 & .... & 22.30\\
  GMRT022538-043420 & 02 25 38.45 & -04 34 17.1  & 2.043$\pm$0.135 & 0.474$\pm$0.015 & -1.00$\pm$0.05 & 0.271 & .... & 23.02\\
  GMRT022539-045706 & 02 25 39.03 & -04 57 09.7  & 1.137$\pm$0.217 & 0.226$\pm$0.024 & -1.11$\pm$0.15 & 2.090 & .... & 24.92\\
  GMRT022539-042823 & 02 25 40.10 & -04 28 21.8  & 2.706$\pm$0.201 & 0.448$\pm$0.052 & -1.23$\pm$0.09 & 0.247 &0.204& 22.74\\
  GMRT022539-041757 & 02 25 40.11 & -04 17 56.0  & 14.37$\pm$0.20  & 2.676$\pm$0.022 & -1.15$\pm$0.01 & 0.779 &0.768& 24.90\\
  GMRT022544-041101 & 02 25 44.65 & -04 10 58.0  & 0.862$\pm$0.110 & 0.195$\pm$0.015 & -1.02$\pm$0.10 & 2.555 & .... & 25.03\\
  GMRT022544-040649 & 02 25 44.96 & -04 06 46.7  & 1.437$\pm$0.154 & 0.206$\pm$0.015 & -1.33$\pm$0.09 & 2.959 &2.643& 25.27\\
  GMRT022545-045857 & 02 25 45.32 & -04 58 54.1  & 2.615$\pm$0.194 & 0.387$\pm$0.016 & -1.31$\pm$0.06 & 2.496 & .... & 25.46\\
  GMRT022550-042141 & 02 25 50.67 & -04 21 41.3  & 0.778$\pm$0.110 & 0.135$\pm$0.016 & -1.20$\pm$0.13 & 3.493 &3.86 & 25.43\\
 $^{\star}$GMRT022556-043523 & 02 25 56.53 & -04 35 23.1  & 0.691$\pm$0.095 & 0.056$\pm$0.018 & -1.72$\pm$0.24 & .... & .... & .... \\
  GMRT022559-041553 & 02 25 59.29 & -04 15 51.7  & 0.552$\pm$0.121 & 0.123$\pm$0.017 & -1.03$\pm$0.18 & 1.162 & .... & 23.98\\
  GMRT022600-041426 & 02 26 01.00 & -04 14 24.4  & 0.601$\pm$0.117 & 0.112$\pm$0.016 & -1.15$\pm$0.16 & 1.187 & .... & 24.01\\
  GMRT022603-042932 & 02 26 03.10 & -04 29 29.2  & 8.067$\pm$0.143 & 1.346$\pm$0.018 & -1.23$\pm$0.02 & 1.599 & .... & 25.45\\
  GMRT022606-045614 & 02 26 06.34 & -04 56 09.8  & 1.316$\pm$0.185 & 0.187$\pm$0.026 & -1.34$\pm$0.13 & 1.950 & .... & 24.87\\
  GMRT022607-044213 & 02 26 07.76 & -04 42 10.4  & 2.311$\pm$0.129 & 0.497$\pm$0.022 & -1.05$\pm$0.05 & 0.271 & .... & 23.05\\
  GMRT022609-040433 & 02 26 10.00 & -04 04 31.2  & 0.976$\pm$0.130 & 0.195$\pm$0.015 & -1.10$\pm$0.10 & 2.787 & .... & 25.17\\
  GMRT022615-044305 & 02 26 15.18 & -04 43 03.8  & 0.646$\pm$0.099 & 0.123$\pm$0.017 & -1.14$\pm$0.14 & 0.156 & .... & 21.91\\
 $^{\star}$GMRT022614-044249 & 02 26 15.19 & -04 43 03.0  & 0.520$\pm$0.095 & 0.107$\pm$0.017 & -1.08$\pm$0.16 & .... & .... & .... \\
  GMRT022619-043050 & 02 26 19.32 & -04 30 47.3  & 1.368$\pm$0.133 & 0.262$\pm$0.015 & -1.13$\pm$0.08 & .... & .... & .... \\
  GMRT022620-042930 & 02 26 20.94 & -04 29 28.6  & 1.356$\pm$0.122 & 0.294$\pm$0.019 & -1.05$\pm$0.08 & 1.151 & .... & 24.36\\
  GMRT022621-040839 & 02 26 21.20 & -04 08 34.7  & 1.095$\pm$0.237 & 0.140$\pm$0.017 & -1.41$\pm$0.17 & 1.764 & .... & 24.66\\
  GMRT022623-041255 & 02 26 23.63 & -04 12 53.0  & 1.473$\pm$0.169 & 0.244$\pm$0.017 & -1.23$\pm$0.09 & 0.299 &0.32 & 22.93\\
  GMRT022624-044205 & 02 26 24.80 & -04 42 03.3  & 3.440$\pm$0.198 & 0.209$\pm$0.017 & -1.92$\pm$0.07 & 1.101 &1.059& 24.39\\
  GMRT022626-041639 & 02 26 27.33 & -04 16 41.5  & 1.266$\pm$0.268 & 0.146$\pm$0.016 & -1.48$\pm$0.16 & 2.932 & .... & 25.33\\
  GMRT022628-044734 & 02 26 28.83 & -04 47 32.2  & 0.576$\pm$0.115 & 0.105$\pm$0.016 & -1.17$\pm$0.17 & 0.300 & .... & 22.49\\
  GMRT022630-045436 & 02 26 30.14 & -04 54 32.6  & 3.612$\pm$0.147 & 0.751$\pm$0.022 & -1.08$\pm$0.03 & 2.391 & .... & 25.58\\
  GMRT022630-043258 & 02 26 30.29 & -04 32 57.7  & 0.583$\pm$0.114 & 0.098$\pm$0.015 & -1.22$\pm$0.17 & 0.107 & .... & 21.46\\
  GMRT022630-045902 & 02 26 30.90 & -04 58 59.3  & 0.635$\pm$0.137 & 0.134$\pm$0.020 & -1.07$\pm$0.18 & 1.906 & .... & 24.58\\
  GMRT022631-043926 & 02 26 31.04 & -04 39 28.8  & 9.111$\pm$0.376 & 1.570$\pm$0.037 & -1.20$\pm$0.03 & 1.425 & .... & 25.38\\
  GMRT022631-042456 & 02 26 31.12 & -04 24 53.3  & 4.557$\pm$0.328 & 0.699$\pm$0.066 & -1.28$\pm$0.08 & ....  & .... &  .... \\
  GMRT022631-044907 & 02 26 31.90 & -04 49 03.3  & 1.043$\pm$0.159 & 0.132$\pm$0.016 & -1.42$\pm$0.13 & 0.653 & .... & 23.47\\
  GMRT022632-044308 & 02 26 32.55 & -04 43 06.2  & 2.082$\pm$0.205 & 0.297$\pm$0.018 & -1.33$\pm$0.08 & 2.845 & .... & 25.51\\
  GMRT022633-045338 & 02 26 33.62 & -04 53 36.5  & 1.557$\pm$0.124 & 0.311$\pm$0.017 & -1.10$\pm$0.07 & 1.928 & .... & 24.97\\
\hline
\end{tabular}
\label{tab:USSSample} 
\end{minipage}
\end{table*}

\addtocounter{table}{-1}

\begin{table*}
\centering
\begin{minipage}{140mm}
\caption{USS sample}
\small
\begin{tabular}{@{}ccccccccc@{}}
\hline
Source    & RA$_{\rm 1.4~GHz}$ &  DEC$_{\rm 1.4~GHz}$ &   S$_{\rm 325~MHz}$ & S$_{\rm 1.4~GHz}$ &  ${\alpha}_{\rm 325 MHz}^{\rm 1.4 GHz}$ & z$_{\rm phot}$ & z$_{\rm spec}$ & logL$_{\rm 1.4~GHz}$  \\
 name     & (hms) &  (dms) &   (mJy)             &  (mJy)            &                                         &                &                &   (W Hz$^{-1}$)       \\ \hline
  GMRT022633-040521 & 02 26 33.76 & -04 05 18.3  & 0.619$\pm$0.139 & 0.105$\pm$0.016 & -1.21$\pm$0.18 & 0.294 & .... & 22.48\\
  GMRT022634-041358 & 02 26 34.27 & -04 13 57.0  & 2.273$\pm$0.152 & 0.412$\pm$0.018 & -1.17$\pm$0.05 & 2.339 & .... & 25.34\\
  GMRT022635-041127 & 02 26 35.11 & -04 11 25.8  & 4.098$\pm$0.128 & 0.805$\pm$0.019 & -1.11$\pm$0.03 & 1.529 & .... & 25.13\\
  GMRT022635-043227 & 02 26 35.85 & -04 32 27.3  & 4.207$\pm$0.153 & 0.840$\pm$0.017 & -1.10$\pm$0.03 & 0.654 &0.694& 24.27\\
  GMRT022636-041644 & 02 26 36.13 & -04 16 41.9  & 2.606$\pm$0.278 & 0.240$\pm$0.019 & -1.63$\pm$0.09 & 1.535 & .... & 24.82\\
  GMRT022637-041247 & 02 26 37.89 & -04 12 46.1  & 9.460$\pm$0.366 & 1.974$\pm$0.024 & -1.07$\pm$0.03 & 0.300 & .... & 23.76\\
  GMRT022640-044607 & 02 26 40.52 & -04 46 07.0  & 2.381$\pm$0.151 & 0.526$\pm$0.022 & -1.03$\pm$0.05 & 1.737 & .... & 25.05\\
  GMRT022641-041800 & 02 26 41.68 & -04 17 55.8  & 1.686$\pm$0.188 & 0.348$\pm$0.019 & -1.08$\pm$0.08 & 1.087 & .... & 24.38\\
  GMRT022642-044625 & 02 26 42.22 & -04 46 25.3  & 15.38$\pm$0.30  & 3.509$\pm$0.033 & -1.01$\pm$0.01 & ....  & .... & .... \\
  GMRT022642-044209 & 02 26 42.78 & -04 42 05.7  & 3.737$\pm$0.200 & 0.220$\pm$0.017 & -1.94$\pm$0.06 & 1.820 & .... & 25.13\\
  GMRT022643-042727 & 02 26 43.67 & -04 27 27.8  & 0.515$\pm$0.110 & 0.090$\pm$0.016 & -1.19$\pm$0.19 & 1.299 & .... & 24.03\\
  GMRT022643-040426 & 02 26 43.80 & -04 04 23.9  & 0.862$\pm$0.132 & 0.135$\pm$0.017 & -1.27$\pm$0.13 & 2.764 & .... & 25.10\\
  GMRT022644-040811 & 02 26 44.64 & -04 08 07.3  & 1.029$\pm$0.155 & 0.200$\pm$0.014 & -1.12$\pm$0.11 & 0.817 & .... & 23.83\\
  GMRT022644-041111 & 02 26 45.02 & -04 11 09.9  & 0.617$\pm$0.140 & 0.123$\pm$0.016 & -1.10$\pm$0.18 & 2.464 & .... & 24.84\\
  GMRT022648-042748 & 02 26 48.36 & -04 27 50.2  & 2.227$\pm$0.251 & 0.349$\pm$0.020 & -1.27$\pm$0.09 & 0.288 &0.328& 23.12\\
  GMRT022656-040328 & 02 26 56.29 & -04 03 26.3  & 3.203$\pm$0.212 & 0.590$\pm$0.017 & -1.16$\pm$0.05 & 3.014 & .... & 25.77\\
  GMRT022656-042234 & 02 26 56.91 & -04 22 32.7  & 0.778$\pm$0.117 & 0.139$\pm$0.017 & -1.18$\pm$0.13 & 2.328 & .... & 24.87\\
  GMRT022658-041815 & 02 26 58.10 & -04 18 14.9  & 2.724$\pm$0.233 & 0.217$\pm$0.016 & -1.73$\pm$0.08 & ....  & .... & .... \\
  GMRT022658-043527 & 02 26 58.99 & -04 35 26.5  & 11.56$\pm$0.66  & 1.658$\pm$0.044 & -1.33$\pm$0.04 & 0.208 & .... & 23.34\\
  GMRT022659-040728 & 02 26 59.69 & -04 07 27.0  & 1.416$\pm$0.133 & 0.276$\pm$0.015 & -1.12$\pm$0.07 & ....  & .... & .... \\
  GMRT022701-042003 & 02 27 00.75 & -04 20 05.8  & 1.951$\pm$0.201 & 0.202$\pm$0.018 & -1.55$\pm$0.09 & 0.347 & .... & 22.98\\
  GMRT022709-042345 & 02 27 09.90 & -04 23 44.8  & 1.513$\pm$0.120 & 0.238$\pm$0.016 & -1.27$\pm$0.07 & ....  & .... & .... \\
  GMRT022712-042412 & 02 27 12.61 & -04 24 11.8  & 1.101$\pm$0.142 & 0.187$\pm$0.016 & -1.21$\pm$0.11 & 1.574 & .... & 24.57\\
  GMRT022718-044319 & 02 27 19.01 & -04 43 21.4  & 1.838$\pm$0.208 & 0.388$\pm$0.017 & -1.07$\pm$0.08 & 0.945 &0.959& 24.28\\
  GMRT022719-041406 & 02 27 19.60 & -04 14 06.4  & 2.826$\pm$0.256 & 0.331$\pm$0.014 & -1.47$\pm$0.07 & 1.109 & .... & 24.50\\
  GMRT022724-042506 & 02 27 24.33 & -04 25 02.2  & 0.908$\pm$0.132 & 0.115$\pm$0.016 & -1.41$\pm$0.14 & 1.715 & .... & 24.54\\
  GMRT022727-040043 & 02 27 27.78 & -04 00 45.0  & 1.789$\pm$0.312 & 0.354$\pm$0.021 & -1.11$\pm$0.13 & ....  & .... & .... \\
  GMRT022727-043735 & 02 27 27.92 & -04 37 34.2  & 77.99$\pm$0.22  & 17.77$\pm$0.02  & -1.01$\pm$0.01 & 1.342 &1.062& 26.04\\
$^{\dagger}$GMRT022728-040344 & 02 27 28.22 & -04 03 42.7  & 24.04$\pm$0.37 & 5.433$\pm$0.025 & -1.02$\pm$0.01 & .... & .... & .... \\
  GMRT022730-041119 & 02 27 30.52 & -04 11 17.8  & 9.255$\pm$0.181 & 1.805$\pm$0.019 & -1.12$\pm$0.02 & 2.197 & .... & 25.89\\
  GMRT022732-044956 & 02 27 31.86 & -04 49 58.3  & 3.805$\pm$0.623 & 0.821$\pm$0.043 & -1.05$\pm$0.12 & .... & .... & .... \\
  GMRT022733-041211 & 02 27 33.37 & -04 12 08.8  & 2.163$\pm$0.195 & 0.391$\pm$0.016 & -1.17$\pm$0.07 & .... & .... & .... \\
$^{\dagger}$GMRT022733-043317 & 02 27 33.61 & -04 33 15.9  & 14.85$\pm$0.35  & 3.109$\pm$0.033 & -1.07$\pm$0.02 & .... & .... & .... \\
  GMRT022735-043201 & 02 27 35.50 & -04 31 59.8  & 0.818$\pm$0.134 & 0.155$\pm$0.016 & -1.14$\pm$0.13 & .... & .... & .... \\
$^{\dagger}$GMRT022735-041121 & 02 27 35.80 & -04 11 22.3  & 64.62$\pm$0.70  & 13.00$\pm$0.030 & -1.10$\pm$0.01 & .... & ....& .... \\
  GMRT022736-040550 & 02 27 35.96 & -04 05 49.7  & 2.169$\pm$0.467 & 0.220$\pm$0.015 & -1.57$\pm$0.15 & 3.101 & .... & 25.63\\
$^{\dagger}$GMRT022743-042130 & 02 27 43.23 & -04 21 28.1  & 12.62$\pm$1.05  & 2.247$\pm$0.077 & -1.18$\pm$0.06 & .... & .... & .... \\
  GMRT022743-043541 & 02 27 43.54 & -04 35 38.9  & 1.255$\pm$0.139 & 0.214$\pm$0.016 & -1.21$\pm$0.09 & 0.334 & .... & 22.92\\
  GMRT022754-044455 & 02 27 54.09 & -04 44 53.8  & 51.13$\pm$0.47  & 10.32$\pm$0.03  & -1.10$\pm$0.01 & ....  & .... & .... \\
  GMRT022757-040749 & 02 27 58.16 & -04 07 45.1  & 4.255$\pm$0.440 & 0.968$\pm$0.022 & -1.01$\pm$0.07 & 0.279 & .... & 23.37\\
\hline
\end{tabular}
\label{tab:USSSample} 
\end{minipage}
\end{table*}

\end{appendix} 

\bibliographystyle{aa}
\bibliography{RadioXMMLSS}
\end{document}